\documentclass{aa}
\usepackage{natbib,psfig,graphicx,ulem}
\usepackage{txfonts}
\usepackage{soul,color}

\def\phzs{photo-{\it z}s\/}
\def\phz{photo-{\it z\/}}

\def\mathnew{\mathsurround=0pt}
\def\simov#1#2{\lower .5pt\vbox{\baselineskip0pt \lineskip-.5pt
\ialign{$\mathnew#1\hfil##\hfil$\crcr#2\crcr\sim\crcr}}}
\def\simgreat{\mathrel{\mathpalette\simov >}}

\def\arcsec{\hbox{$^{\prime\prime}$}}

\def\MeV{Me\kern-0.11em V}
\def\keV{ke\kern-0.11em V}
\def\arcsec{\hbox{$^{\prime\prime}$}}

\begin{document}         
\sethlcolor{red} \title{The DAFT/FADA survey. I.~Photometric redshifts
  along lines of sight to clusters in the z=[0.4,0.9] interval
  \thanks{Based on observations made with the NASA/ESA Hubble Space
    Telescope, obtained from the data archive at the Space Telescope
    Institute and the Space Telescope European Coordinating
    Facility. STScI is operated by the association of Universities for
    Research in Astronomy, Inc. under the NASA contract NAS
    5-26555. Also based on observations made with ESO Telescopes at
    Paranal and La Silla Observatories under programme ESO LP
    166.A-0162. Also based on  visiting astronomer observations, at
    Cerro Tololo Inter-American Observatory, National Optical
    Astronomy Observatory, which is operated by the Association of
    Universities for Research in Astronomy, under contract with the
    National Science Foundation.}}

\author{
L.~Guennou\inst{1,2} \and
C.~Adami\inst{1} \and
M.P.~Ulmer\inst{2,1} \and
V.~LeBrun\inst{1} \and
F.~Durret\inst{3,4} \and
D.~Johnston\inst{6} \and
O.~Ilbert\inst{1} \and
D.~Clowe\inst{5,13} \and
R.~Gavazzi\inst{3,4} \and
K.~Murphy\inst{5} \and
T.~Schrabback\inst{10} \and \\
S.~Allam\inst{6} \and
J.~Annis\inst{6} \and
%S.~Arnouts\inst{1} \and
S.~Basa\inst{1} \and
C.~Benoist\inst{7} \and
A.~Biviano\inst{8} \and
A.~Cappi\inst{9} \and
%J.G.~Cuby\inst{1} \and
%C.~Ferrari\inst{7} \and
%J.~Hao\inst{6} \and
%J.P.~Kneib\inst{1} \and
%R.~Kron\inst{10} \and
J.M.~Kubo\inst{6} \and
%H.~Lin\inst{6} \and
%R.~Malina\inst{1} \and
P.~Marshall\inst{11,12} \and
%S.~Maurogordato\inst{7} \and
A.~Mazure\inst{1} \and
F.~Rostagni\inst{7} \and
D.~Russeil\inst{1} \and
%C.~Schimd\inst{1} \and
E.~Slezak\inst{7}
%D.~Tucker\inst{6} 
}

\institute{
LAM, OAMP, P\^ole de l'Etoile Site de Ch\^ateau-Gombert,
38 rue Fr\'ed\'eric Joliot-Curie,
13388 Marseille Cedex 13, France
\and
Department Physics $\&$ Astronomy, Northwestern University, Evanston, 
IL 60208-2900, USA
\and
UPMC Universit\'e Paris 06, UMR~7095, Institut d'Astrophysique de Paris, 
F-75014, Paris, France
\and
CNRS, UMR~7095, Institut d'Astrophysique de Paris, F-75014, Paris, France
\and
Department of Physics and Astronomy, Ohio University, 251B Clippinger
Lab, Athens, OH 45701, USA
\and
Fermi National Accelerator Laboratory, P.O. Box 500, Batavia, IL 60510, USA 
\and
OCA, Cassiop\'ee, Boulevard de l'Observatoire, BP 4229, 06304 Nice Cedex 4, France
\and
INAF/Osservatorio Astronomico di Trieste, via G. B. Tiepolo 11, I-34143, 
Trieste, Italy 
\and
INAF - Osservatorio Astronomico di Bologna, via Ranzani 1, 40127 Bologna, Italy
%\and
%University of Chicago, Department of Astronomy and Astrophysics, 5640 South Ellis Avenue, Chicago, IL 60637, USA
\and
Leiden Observatory, Leiden University, Niels Bohrweg 2, NL-2333 CA
Leiden, The Netherlands
\and
Kavli Institute for Particle Astrophysics and Cosmology, Stanford University, 
2575 Sand Hill Road, Menlo Park, CA 94025, USA
\and
Physics Department, University of California, Santa Barbara, CA 93601, USA
\and
Alfred P. Sloan Fellow
}

\date{Accepted . Received ; Draft printed: \today}

\authorrunning{Guennou et al.}

\titlerunning{The DAFT/FADA survey. I.~Photometric redshifts.}

\abstract
% context heading (optional)
{ As a contribution to the understanding of the dark energy concept,
  the Dark energy American French Team (DAFT, in French FADA) has
  started a large project to characterize statistically high redshift
  galaxy clusters, infer cosmological constraints from Weak Lensing
  Tomography, and understand biases relevant for constraining dark
  energy and cluster physics in future cluster and cosmological
  experiments. }
% aims heading (mandatory)
{ The purpose of this paper is to establish the basis of
  reference for the photo-$z$ determination used in all our subsequent
  papers, including weak lensing tomography studies.}
% methods heading (mandatory)
{This project is based on a sample of 91 high redshift ($z\geq$0.4),
  massive ($\simgreat 3 \times 10^{14}$~M$_\odot$) clusters with
  existing HST imaging, for which we are presently performing
  complementary multi-wavelength imaging. This allows us in particular
  to estimate spectral types and determine accurate photometric
  redshifts for galaxies  along the lines of sight to the first
    ten clusters for which all the required data are available down to
    a limit of ${\rm I_{AB}}=24./24.5$ with the LePhare software. The
    accuracy in redshift is of the order of 0.05 for the range $0.2
    \leq z \leq 1.5$.}
% results heading (mandatory)
{ We verified that the technique applied to obtain photometric
    redshifts works well by comparing our results to with previous
  works.  In clusters, \phz\ accuracy is degraded for bright absolute
  magnitudes and for the latest and earliest type galaxies. The \phz\
  accuracy also only slightly varies as a function of the spectral
  type for field galaxies.  As a consequence, we find evidence
    for an environmental dependence of the \phz\ accuracy, interpreted
as the standard used Spectral Energy Distributions being not very well suited 
to cluster galaxies. Finally, we
    modeled the LCDCS 0504 mass with the strong arcs detected along
    this line of sight. }
% conclusions
{}

\keywords{}

\maketitle

\section{Introduction}
\label{introduction}

The discovery ten years ago of the acceleration of the expansion of
the Universe (Riess et al. 1998) which is typically explained by
assuming that most of its energy is in the form of an unknown dark
energy (DE), is one of the most puzzling issues of modern
cosmology. Efforts have therefore been undertaken, such as the Dark
Energy Task Force (Albrecht et al.  2006) or the ESA-ESO working group
on fundamental physics (Peacock et al.  2006) to design projects to
measure DE and determine its nature. As highlighted by these reports,
understanding DE requires big surveys to overcome cosmic variance and
shot noise as well as new experiments to control the unknown
systematic uncertainties.

In this context, galaxy clusters, together with several other probes,
are expected to play a major role (e.g. Nichol 2007). These objects
have indeed long held a place of importance in astronomy and
cosmology.  Zwicky (e.g. 1933) inferred from observations of the Coma
cluster that the matter in our universe could be in the form of a
dark component (this component was first supposed to be low surface brightness
diffuse light). The measurement of the baryon fraction in X-ray
clusters (e.g. Lubin et al. 1996, Cruddace et al. 1997), combined with
big-bang nucleosynthesis constraints allowed to put an upper limit on
the matter density of the Universe. The resulting value was
considerably less than the theoretically-favored critical
density. Cluster number counts (e.g. Evrard 1989) and cluster
correlation functions (e.g. Bahcall \& Soneira 1983) have been used to
constrain the amplitude of mass fluctuations and strengthen support
for the Cold Dark Matter (CDM) structure formation paradigm.  Galaxy
clusters can also be used to test the redshift-distance relation
(e.g. Supernovae as standard candles, Baryon Acoustic Oscillations or
Weak Lensing Tomography with clusters, e.g. Hu 1999) or the growth of
structures through weak lensing, cluster number counts, or integrated
Sachs-Wolfe effect. Clusters are also intrinsically interesting in
many aspects, including the influence of environment on galaxy
formation and evolution. Building a detailed picture of galaxy and
large-scale structure growth (e.g. clusters) is therefore necessary to
understand how the Universe has evolved.

The Dark energy American French Team (DAFT, in French FADA) has
started a large project to characterize statistically high redshift
galaxy clusters, infer cosmological constraints from Weak Lensing
Tomography, and understand biases relevant for constraining DE and
cluster physics in future cluster and cosmological experiments. This
work is based on a sample of 91 high redshift (z=[0.4;0.9]),
massive ($>3 \times 10^{14}$~M$_\odot$) clusters with existing HST
imaging, for which we are presently performing complementary
multi-wavelength imaging. This will allow us in particular to estimate
accurate photometric redshifts for as many galaxies as possible.  
  The requested accuracy depends on both our ability to discriminate between
  cluster and background field galaxies without loosing too many
  objects and on the Weak Lensing Tomography method internal parameters. 
  Catalogs of cluster galaxies (e.g. Adami et al. 2008)
  typically show photometric redshifts spanning a total ($\sim
  3\sigma$) interval of $\pm$0.15 in photo$-z$. This means that the
  goal of our survey is to have photometric redshifts with a 1$\sigma$
  precision better than 0.05. With such a precision, the 
photo-$z$ uncertainties would not be the expected dominant source of errors in 
our method,
except when considering lensing and lensed objects at redshift greater than 0.8
and closer than 0.4 along the redshift direction.  The
immediate goal of this paper is then to describe these photometric
redshift measurements on the first 10 completed clusters in our
sample. This will allow us in the future to combine photo-$z$s with
weak lensing shear measurements both to carry out tomography and to
build mass models for clusters.  This paper will also provide the
foundation for other future works that will use the photo-$z$s
produced by the process described here to study cluster galaxy
populations.

Throughout the paper we assume H$_0$ = 71 km s$^{-1}$ Mpc$^{-1}$,
$\Omega _m$=0.27, and $\Omega _{\Lambda}$=0.73. All magnitudes are in
the $AB$ system.

\section{Observations}
\label{reduction}

The 10 clusters for which we produced photometric redshifts (hereafter
\phzs\ ) were originally observed as part of the EDisCS program
(e.g. White et al. 2005), but new data (mostly B, but also R and z$'$,
see Table~\ref{tbl}) were obtained and some data were also collected
from the literature (V, R, I, z$'$, F814W, Spitzer IRAC) to complete
the data set in order to calculate more accurate \phzs. We thus
created a full data set with BVRIz', HST ACS F814W, and Spitzer IRAC
3.6 $\mu$m and 4.5$\mu$m (channels 1 \& 2). Fig.~\ref{lo} shows the
spectral coverage achieved with this set of filters.

\begin{figure}[!h]
  \begin{center}
    \caption{Transmission curves of the available sets of
      filters. Upper figure: infrared filters (Irac 1: cyan, Irac2:
      blue). Lower figure: visible filters (from left to right: B, V,
      R, I in green, F814W in yellow, and z$'$).}
  \label{lo}
  \end{center}
  \end{figure}

\subsection{HST ACS data}

We have retrieved from the HST archives data for 10 EDisCS clusters 
observed with the ACS in
the F814W filter, each image including 4 tiles (2$\times$2 mosaic) of
2~ks and a central tile of 8~ks (Desai et al. 2007). The achieved depth 
for point sources at the 90$\%$ level is of the order of F814W$\sim$28 for the deep parts 
and F814W$\sim$26 for the shallow parts (see Fig.~\ref{comp})
 \footnote{Considering total magnitudes and analysis being performed at the 1.8 and 
the 3 sigma Sextractor level respectively for the deep and shallow parts to limit fake 
object detections.}.
The full data
reduction technique is described in Schrabback et al. (2010) and will
be expanded in a companion paper (Clowe et al. in preparation), but we
give here the salient points.  The data were reduced using a modified
version of the HAGGLeS pipeline, with careful background subtraction,
improved bad pixel masking, and proper image registration.  Stacking
and cosmic ray rejection were done with Multidrizzle (MD) (Koekemoer
et al. 2002), taking the time-dependent field-distortion model from
Anderson et al. (2007) into account.  The pixel scale was 0.05 arcsec
and we used a Lanczos3 kernel. After aligning the exposures of each
tile separately, shifts and rotations between the tiles were
determined from separate stacks by measuring the positions of objects
in the overlap regions.  As final step, mosaic stacks including all
tiles of one cluster were created. We set these ACS mosaics as
astrometric references for ground-based data. Our first results based
on weak lensing measurements and weak lensing tomography will be
described in a companion papers (Clowe et al., in preparation).

\subsection{B, R and z$'$ ground based data}

The new B, R and z' band observations presented here were conducted at
the CTIO Blanco telescope using the multi-CCD device MOSAIC (see
Table~\ref{tbl}). Exposure times were computed
to reach an expected depth of F814W$\sim$24.5 (AB) at the 10$\sigma$
level. The seeing was on average about 1.1~arcsec, 0.7~arcsec and
0.9~arcsec for the B, R and z bands respectively.  During the
observations, we followed a regular dithering pattern with an
amplitude of 5--10~arcsec\ to improve cosmetics (e.g. inter-chip
separations) of the final images. The standard star fields SA98,
SA104, SA107a and SA107b were also regularly observed during the
nights (3 standard stars per night). This
allowed us to derive extinction curves for the Blanco site and perform
photometric calibrations.

\begin{table*}
\caption{Decimal degrees J2000 coordinates, observed bands, CTIO exposure times, and redshift for 
each of the ten considered clusters.}
\label{tbl}
\begin{center}
%\begin{tabular}{|c|c|c|c|c|c|} 
\begin{tabular}{ccccccc} 
\hline
cluster name & RA & DEC & band & exposure time & z & 90$\%$ completeness level\\ 
           & deg & deg &  & seconds &  & F814W magnitude\\ \hline
LCDCS 0110 & 159.464 & $-12.724$  & B & 11x600 & 0.58 & 26.2 \\ \hline
LCDCS 0130 & 160.168  & $-11.934$  & B & 11x600 & 0.70 & 26.2 \\ \hline
LCDCS 0172 & 163.601  & $-11.772$  &  B & 11x600 & 0.70 & 26.2 \\ \hline
LCDCS 0173 & 163.681  & $-12.764$  &  B & 11x600 & 0.75 & 26.2 \\ \hline
CLJ1103.7-1245a & 165.895  & $-12.780$  &  B & 11x600 &  0.63 & 26.2 \\ \hline
LCDCS 0340 & 174.542  & $-11.560$  &  B & 11x600 & 0.48 & 26.2 \\ \hline
LCDCS 0504 & 184.189  & $-12.022$  &  B & 11x600 &  0.79 & 26.2 \\ \hline
LCDCS 0531 & 186.995  & $-11.587$  &  B & 11x600 & 0.64 & 26.2 \\ \hline
LCDCS 0541 & 188.126  & $-12.8434$  &  z' & 18x500 & 0.54 & 25.8 \\ \hline
           &   &   &  R & 8 x 600 & & 26.6 \\ \hline
LCDCS 0853 & 208.541  & $-12.517$  &  B & 11x600 & 0.76 & 26.2 \\ \hline
\end{tabular}
\end{center}
\end{table*}

\subsection{Tools}

For the image reduction, we used the MIDAS, SCAMP and SWarp
(e.g. Bertin et al. 2002, Bertin 2006) packages to produce images with
cosmic rays and other image defects removed, and to produce final
calibrated and aligned images; these newly acquired data have then
been combined with the previously reduced data.  Descriptions of the
MissFits, SCAMP and SWarp software are given in
http://www.astromatic.net.  We co-aligned the FITS images in different
bands by using SCAMP and then combined them to generate panchromatic
images (see e.g. Fig.~\ref{trich}).

\subsection{Ground based data reduction}

%We acquired offsets and flat fields to correct for instrumental
%effects. We subtracted an average of the offsets to the scientific
%images and to the flat fields and divided the images by the normalized
%flat fields. We preferred sky flats instead of dome flats because dome
%flats gave us a less uniform background. Sky flats were observed at
%the beginning and end of each night for a total of 21, 19 and 11
%exposures for the B, z$'$ and R bands respectively. For each band, we
%merged the exposures into one and used a minimal filter to take away
%the apparent stars. Then, we normalized the flat fields to 1 in terms of
%total flux.

After the classical reduction scheme (offsets, flatfields, etc.), 
we realized that the gains between the different MOSAIC CCDs were not 
initially very well constrained. To solve this problem, we observed a SDSS 
field in all the bands, which allowed us to adjust these gains. These 
corrections were in most cases smaller than 10\%.

The SCAMP and SWarp tools were used to perform the astrometry and
homogenize internal photometry in order to put each of the individual
images on a common grid to create a merged image without cosmic rays,
with inter-chip gaps filled and CCD defaults erased. This is a commonly
employed technique for the CFHT Megacam and CFH12K images (e.g.
McCracken et al. 2003). We usually found it necessary to use
a third-order polynomial to model the astrometric distortions. We then
generated weight maps which took into account bad pixels, overscans
and low efficiency areas. Fig.~\ref{terapix2}
shows the resulting astrometric
distortion map obtained for one of the B band observations.

%\begin{figure}[!h]
%  \begin{center}
%    \includegraphics[width=3.00in,angle=270]{fgroups_1_ctio2.ps}
%    \caption{CCD focal plane reconstruction by SCAMP
%      (showing all individual exposures) for one of our CTIO B band
%      observations. Red squares are known literature object
%      positions. Green dots are preliminary detections inside our B
%      band image. The large black rectangles are the individual MOSAIC
%      CCDs.}
%  \label{terapix1}
%  \end{center}
%  \end{figure}

\begin{figure}[!h]
  \begin{center}
    \caption{Resulting astrometric distortion map for one of our B
      band observations. Large black rectangles are the individual MOSAIC
      CCDs. Blue to red colors show 0.25 to 0.27 arcsec astrometry
      residual uncertainties.}
  \label{terapix2}
  \end{center}
  \end{figure}

We then used standard stars observed during the nights 
to build extinction curves. \footnote{The respective
extinction coefficients $K$ are 0.202, 0.095 and 0.04 for the B, R and z'
bands, and the corresponding errors on zero points are 0.09 in B and
0.07 mag in R and z'. Observed magnitudes have then to be diminished by
$K$ times the airmass.}

The CCDs from MOSAIC are affected by some cross talk. We corrected for
this effect by subtracting from a contaminated (receiving) CCD the
contaminating (sending) CCD weighted by a factor given on the web page
of MOSAIC (http://www.lsstmail.org/noao/mosaic/calibs.html). In any
case, this only affected bright magnitude objects. However, considering
that the number of such objects per CCD was non-negligible, we had to
take cross-talk into account.

\begin{figure}[!h]
  \begin{center}
    \caption{LCDCS 0541 tricolor image (50 arcsec $\times$ 30 arcsec) 
      made with the B (CTIO: shown
      as blue), F814 (HST: shown as green) and z$'$ (VLT: shown as red)
      filters. Known spectroscopic redshifts are also shown.}
  \label{trich}
  \end{center}
  \end{figure}

\subsection{Completeness level}

Although completeness per se is not important for our goal of
obtaining photo-$z$s of the background searched galaxies, it is
interesting to discuss this issue because net completeness determines
how many galaxies we will have in the end to carry out weak lensing
tomography and cluster studies.  The first important parameter to
estimate is the completeness level of our images in each
band. Based on this information we can then determine the magnitude
range where all our bands can contribute.  To estimate this level, we
ran simulations for the HST ACS F814W images as in Adami et
al. (2006). In brief, the simulation method adds 100 artificial
stars of different magnitudes to the CCD images and then attempts to
recover them by running SExtractor again with the same parameters used
for object detection on the original images. In this way, the
completeness is measured on the original images. We investigated the
catalog completeness for point-like sources only. The completeness
levels in magnitudes are therefore an upper value for the real
completeness level for galaxies of different types.

  For the other bands, we computed how many objects detected in the
  HST ACS F814W images also gave a successful magnitude measurement in
  the other bands when extracted in SExtractor double-image mode (to increase
  the depth of the catalogs). We
  note that the HST ACS F814W images are the deepest among all our
  bands by a large factor. Multiplying these percentages (not always 100$\%$
  because objects are sometimes so close to the background in a given band that
  the fluxes are not significantly positive) by the
  completeness of the HST ACS F814W images themselves, we therefore
  estimate the $complete~detection~level$ of the considered band (as
  function of the F814W magnitude). 

  This method also has the following consequence: we must take into
  account the varying exposure times in the HST ACS F814W images (the
  centers have longer exposure times than the edges). This results in
  a varying mean signal to noise in the HST ACS F814W images and we
  had therefore to adapt the extraction and measure SExtractor signal
  to noise ratios.  We therefore estimated the point source completeness
  separately in the deep (10 ksec) and shallow (2 ksec) portions of
  the HST ACS F814W images. This also gave two different
  $complete~detection~levels$ for the other bands.

\begin{figure}[!h]
  \begin{center}
    \includegraphics[angle=270,width=3.00in]{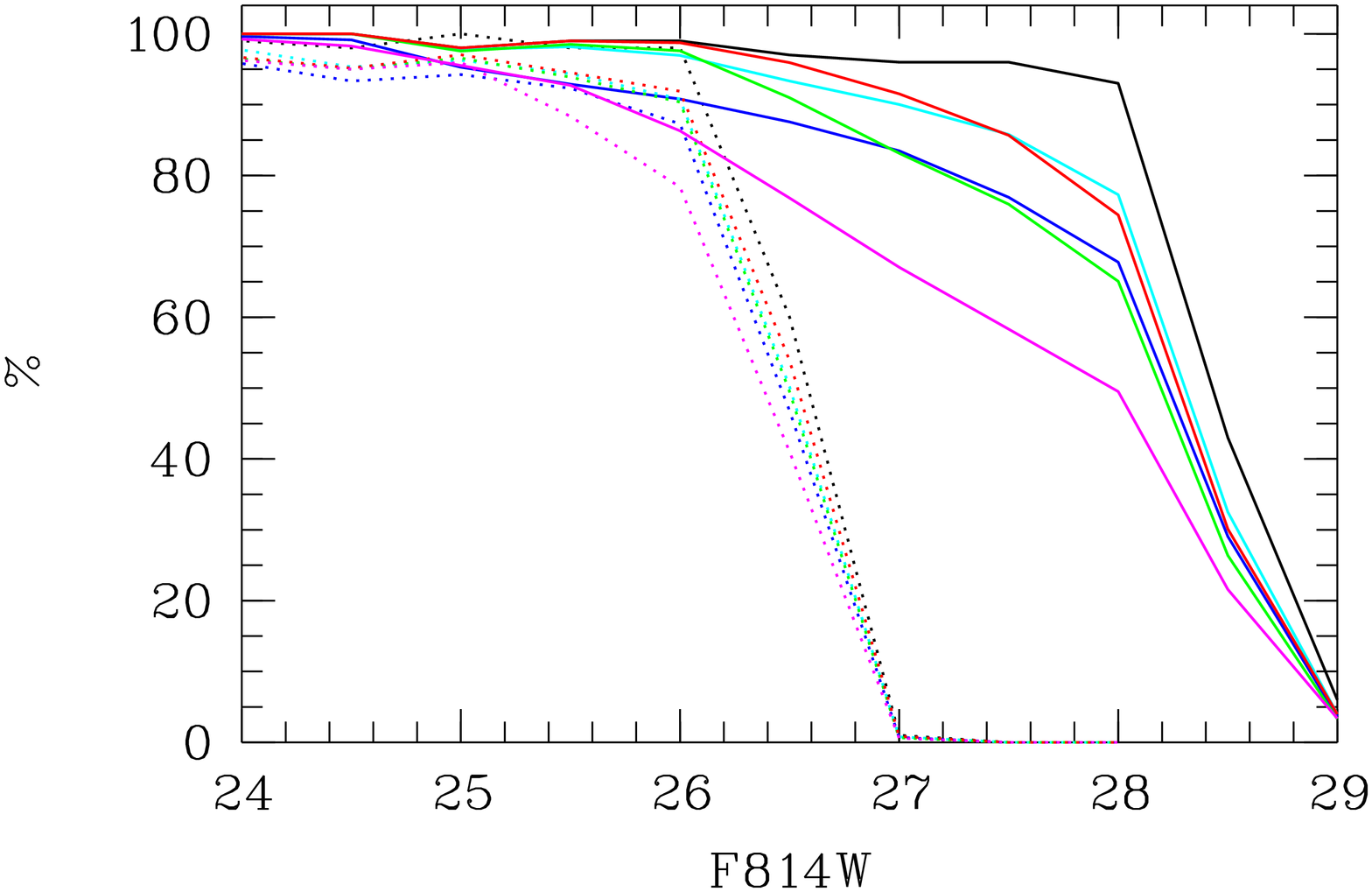}
    \includegraphics[angle=270,width=3.00in]{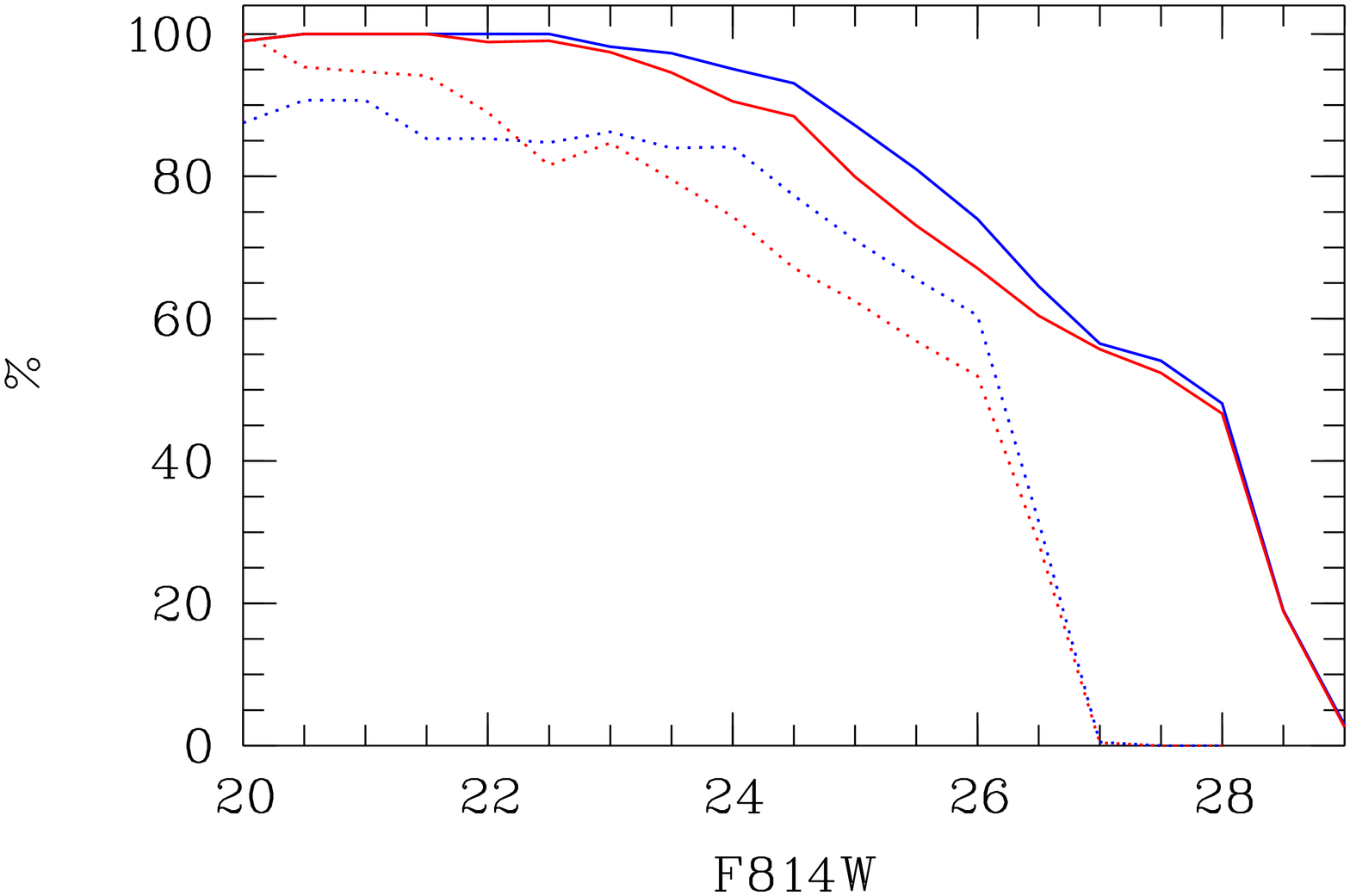}
    \caption{Detection levels for point sources in the LCDCS 0541
      fields. Continuous lines correspond to the deep parts of the HST ACS
      F814W images and dotted lines to the shallow parts. Top
      figure: blue: B band, cyan: V band, green: R band, red: I band,
      black: F814W band, magenta: z' band. Bottom figure: blue: Irac~1
      band, red: Irac~2 band. Each curve is given as a function of the 
      F814W magnitude.}
  \label{comp}
  \end{center}
  \end{figure}

The results are displayed in Fig.~\ref{comp} for LCDCS 0541 (other
clusters show very similar results). This shows that considering point
sources brighter than F814W=24.5 in the HST ACS F814W deep parts
ensures that our data are more than 90\% complete whatever the
considered band. In the HST ACS F814W shallow parts, the same limit
ensures that our data are more than 90\% complete for the ground based
plus HST ACS F814W images, and more than 80\% complete for the Spitzer
data. F814W=24.5 therefore seems to be a reasonable compromise between
depth and completeness level. This value is also close to the expected
value given the chosen exposure times.  If we accept lower
completeness levels of the order of 50\% in all the bands, we can
consider magnitudes as faint as F814W=26. Since we can carry out shear
measurements down to even fainter magnitudes, this completeness result
means we will be able to use a sizable fraction of those shear
measurements.

\sethlcolor{yellow}

\begin{figure}[!h]
  \begin{center}
    \includegraphics[angle=270,width=3.00in]{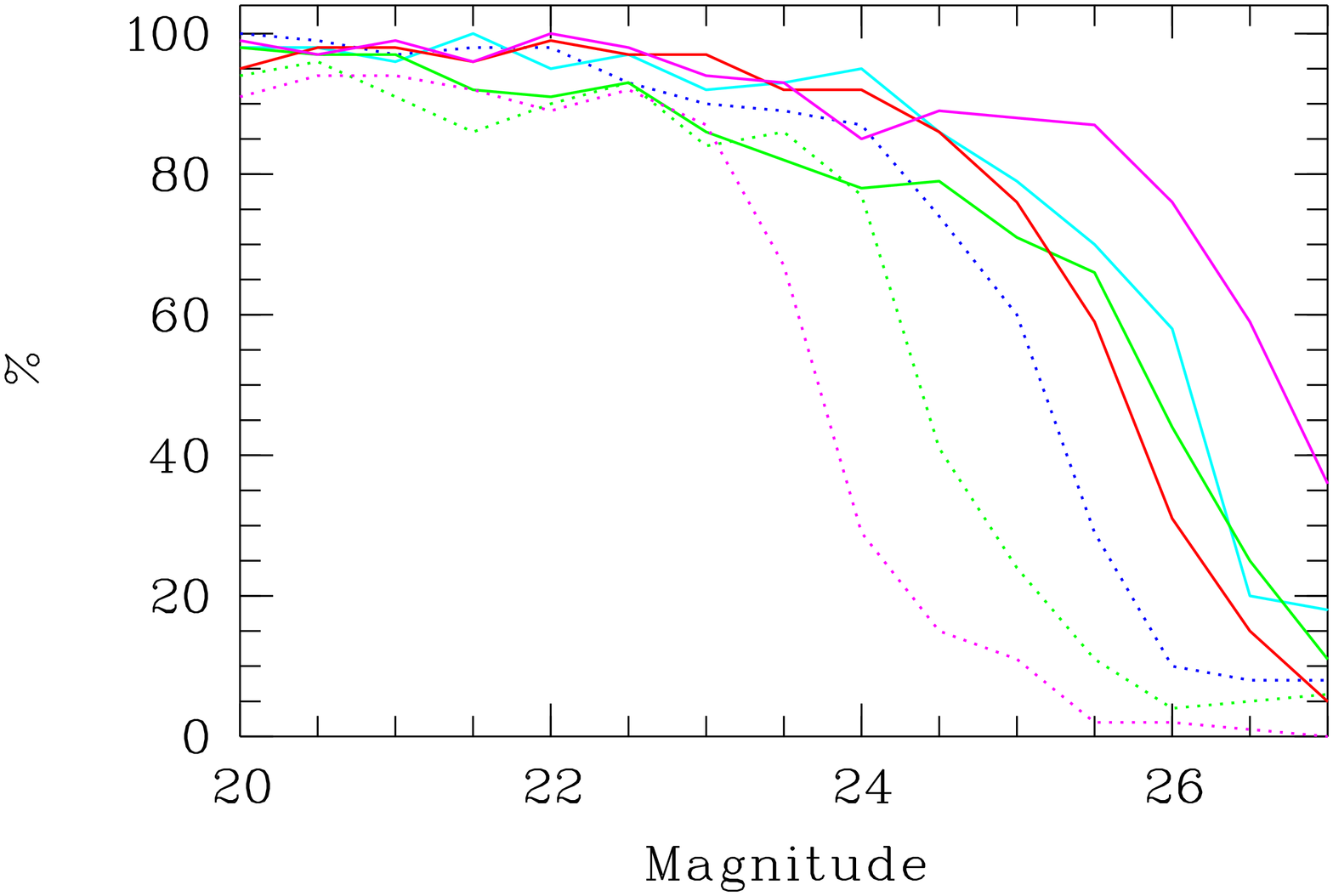}
    \includegraphics[angle=270,width=3.00in]{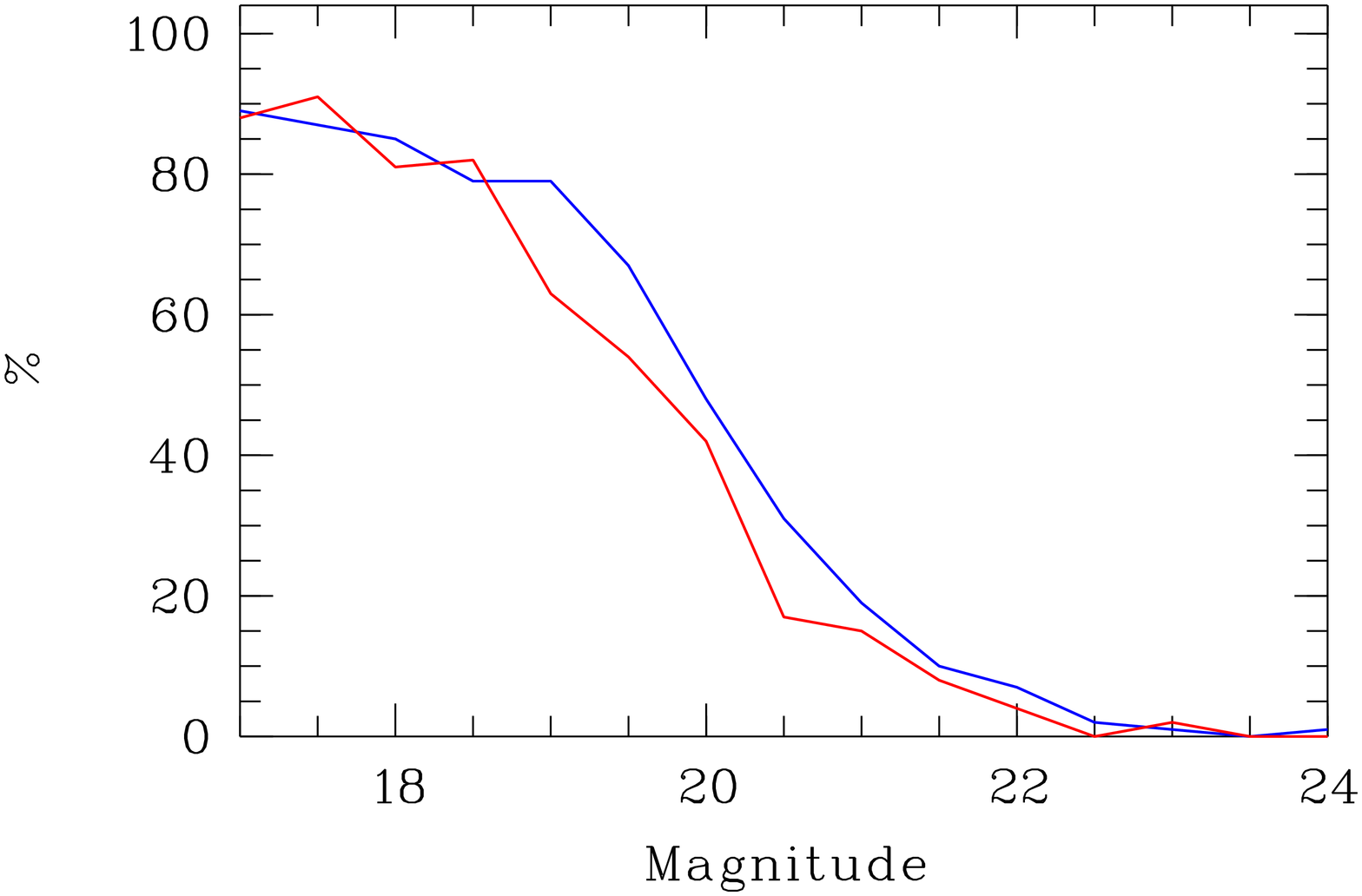}
    \caption{Top figure: detection levels for point sources in the B, V, 
      R, I, z' bands. Continuous lines correspond to the EDISCS data and 
      dashed lines correspond to the CTIO data. blue: B band, cyan: V band, 
      green: R band, red: I band, magenta: z' band. Bottom figure: blue: Irac~1
      band, red: Irac~2 band.}
  \label{compgen}
  \end{center}
  \end{figure}

$Detection~levels$ are also interresting to compute for future uses for a 
given band in itself, detecting $and$ measuring the fluxes in this band 
without considering the HST ACS F814W image as a detection image. We note 
however that resulting catalogs would then be shallower than previously. We 
therefore
performed the same simulations we did for the HST ACS F814W images for the
B, V, R, I, z', Irac~1, and Irac~2 bands (Sextractor detection level of 1.8,
point sources). Fig.~\ref{compgen} gives the corresponding results.  

\sethlcolor{yellow}

\subsection{Star/galaxy separation}

The second step in our processing was to carry out a star-galaxy
separation. We only applied this task to the HST ACS F814W images
because they have by far the best seeing.  For this task we employed
the classical method consisting in separating stars from galaxies (and
from defects) in central surface brightness versus total magnitude
plots. We show in Fig.~\ref{stargal} that the star/galaxy
discrimination is efficient up to F814W$\sim$27 (fainter
than the initially expected depth for our multi wavelength survey).

\begin{figure}[!h]
  \begin{center}
    \caption{Central surface brightness versus total magnitude for the
      HST ACS F814W LCDCS0541 image.  Blue dots are considered as
      galaxies, red dots as stars, and green dots as defects.}
  \label{stargal}
  \end{center}
  \end{figure}

\begin{figure}[!h]
  \begin{center}
    \includegraphics[angle=270,width=3.00in]{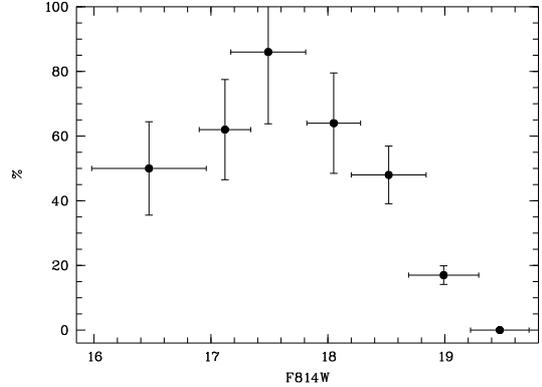}
    \caption{Percentage of saturated stars classified as galaxies by
      our method as a function of magnitude (see text).}
  \label{badstar}
  \end{center}
  \end{figure}

  We also had to deal with bright saturated stars. These objects were
  usually classified as galaxies by classical star/galaxy separation
  methods. Since we were mainly interested in faint objects, these
  saturated stars were not really a problem in our analysis.  For
  completeness of our description of our work, however, we
  investigated what was the contribution of these objects. We examined
  by eye all F814W$\leq$20 LCDCS 0541 objects and determined a real
  star catalog. From this, we determined which objects in this list
  were classified as stars by our automated method based on
  Fig.~\ref{stargal} and we then deduced the percentage of stars
  incorrectly classified as galaxies. Fig.~\ref{badstar} gives these
  percentages as a function of magnitude. We clearly see that limiting
  our analysis to F814W$\geq$19.5 probably ensures that our galaxy
  catalogs are not polluted by stars.

\section{Registration with previous images and resulting \phz\ computation}

We now describe how our new data were combined with the previously
acquired data, and how the \phzs\ were calculated. We also discuss
the reliability of these \phzs.

\subsection{General strategy}

In order to calculate \phzs, the data set had to be combined with data
available in the literature.  We first re-sampled with SCAMP and SWarp
all images to the pixel size of the HST images and the image
astrometry was similarly homogenized to produce an alignment precision
of the order of one pixel between different bands (1~pixel=0.05~arcsec
after realignment).

At this step, the very large CTIO Blanco MOSAIC images
($\sim$30'$\times$30'), the VLT FORS2 images, and the Spitzer IRAC
images were truncated to the size of the HST ACS F814W mosaics.

We homogeneously detected and measured objects with the SExtractor
package in double image mode (Bertin $\&$ Arnouts 1996). Detection was
made on the deepest image (HST ACS F814W image), and measurements were
made inside the HST Kron apertures on the ground based B, V, R, I,
z$'$ images and on the Spitzer Irac~1 and Irac~2 images. At this step,
we only kept objects detected in all the available bands.

\begin{figure}[!h]
  \begin{center}
    \includegraphics[angle=270,width=3.00in]{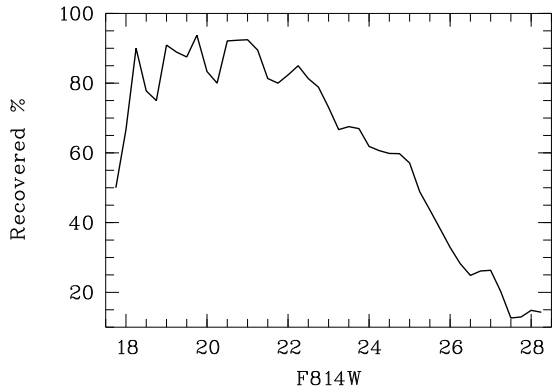}
    \caption{Percentage of recovered objects (detection threshold of 1.8 in 
      SExtractor) in the HST ACS LCDCS 0541
      image when degrading the image to a 1 arcsec resolution, as a
      function of the F814W magnitude.}
  \label{lost}
  \end{center}
\end{figure}

We could have degraded the HST images to the resolution of the ground
based images before extracting sources.  However, because we are
dealing with clusters of galaxies, this would have resulted in the
loss of significant numbers of galaxies in the dense
cores. Fig.~\ref{lost} shows the percentage of recovered objects when
degrading the LCDCS 0541 F814W image to a 1~arcsec resolution. We
clearly see that $\sim$40 \% of the objects are lost at a magnitude of
F814W=24.5 and this would strongly penalize our survey.

\subsection{Image resolution}

Detecting objects in HST images with a spatial resolution sometimes
$\sim$10 times better than other images also has non negligible
consequences. For example, total magnitudes measured in ground based
or in Spitzer images are probably not correct estimates: fluxes
computed from $\sim$1~arcsec spatial resolution images inside the HST
Kron aperture are obviously underestimates of the total
fluxes. However, the main goal of this measure is not to estimate the
total magnitude of the objects but to compute \phzs. The problem can
therefore be solved by using the LePhare \phz\ package (e.g. Ilbert et
al. 2006). Briefly, the LePhare package is able to compare observed
magnitudes with predicted ones created by templates from the
literature as for example in HyperZ (Bolzonella et al., 2000) or
e.g. in Rudnick et al. (2003). We selected spectral energy
distributions (hereafter SEDs) from Polletta et al. (2006, 2007) with
a Calzetti et al. extinction law (e.g.  Calzetti $\&$ Heckman
1999). These are the SEDs which give the best results.  The fitting
then allows to constrain simultaneously the redshift and nature of
each object (galaxy or star), as well as its characteristics such as
photometric type (hereafter $T$).  With the selected class of SED, $T$
varies from 1 to 31. Numbers between 1 and 7 correspond to elliptical
galaxies, numbers between 8 and 12 to S0, Sa, and Sb galaxies (early
spiral galaxies), numbers between 13 and 19 to Sc, Sd, and Sdm
galaxies (late spiral galaxies), and numbers between 20 and 31 to
active galaxies.  LePhare is also able to estimate possible shifts in
photometric values, by comparing the photometric and spectroscopic
redshifts used for training sets, and all the clusters considered in
this paper have deep spectroscopic catalogs (see Fig.~\ref{range}) of
$\sim$100 redshifts per line of sight (Halliday et al. 2004 and
Milvang-Jensen et al. 2008). Shifts are computed fixing \phzs\ to the
spectroscopic values and averaging the residuals in each of the bands.

This is useful to take into account internal photometry
inhomogeneities between different bands, and also allows us to take
into account the different spatial resolutions with LePhare by
applying zero point shifts to our magnitudes.  The mean (over the ten
clusters) applied zero point shifts before \phz\ computations in the
present paper are $0.00\pm$0.14 (B band), $-0.22 \pm0.06$ (V), $-0.20
\pm 0.07$ (R), $-0.18 \pm 0.10$ (I), $-0.34 \pm 0.08$ (F814W),
0.37$\pm$0.11 (z$'$), $-0.86 \pm 0.30$ (Irac~1), $-0.92 \pm 0.31$
(Irac~2).  These values are not negligible, mainly for the Irac~1 and
Irac~2 bands. We will now try to estimate the relative contributions
to these values of the zero point shifts and of the shifts due to the
different spatial resolutions.

A first test to evaluate the influence of the spatial resolution is to
compute the shifts to apply in LePhare when degrading the F814W images
to a 1~arcsec resolution (close to the resolution of the ground based
images). In this case, only the Irac1 and 2 images should require
large shifts because they have the worst resolution, of the order of 2
arcsec. The values are equal (for LCDCS 0541) to 0.01 (B band),
$-0.14$ (V), $-0.18 $ (R), $-0.03$ (I), $-0.15$ (F814W), 0.12 (z$'$),
$-0.39$ (Irac~1), $-0.44$ (Irac~2). In this case, shifts are rather
small and the only large values occur for the two Irac bands, as
expected. This means that a large part of the shifts applied to
magnitudes in LePhare is probably due to different image resolutions.

\begin{figure}[!h]
  \begin{center}
    \includegraphics[angle=270,width=3.00in]{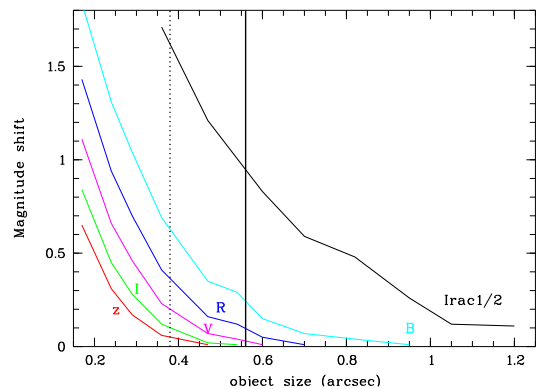}
    \caption{Simulated magnitude shifts to apply in LePhare as a
      function of object size and for the various bands. The solid
      vertical line is the median value of the object sizes for
      F814W$\leq$24.5 in real data. The vertical dashed line shows the
      minimal size of 90$\%$ of the objects.}
  \label{shiftres}
  \end{center}
\end{figure}

We can also evaluate the effect of different resolutions on the images
with simulations.  We generated artificial objects with FWHM varying
from to 0.2 to 1.2 arsec and applied the magnitude measurement process
to these images as seen with the HST F814W configuration and as seen
in the other bands. Namely, these are B band (seeing of 1.05~arcsec),
V (0.68~arcsec), R (0.82~arcsec), I (0.62~arcsec), z' (0.54~arcsec),
and Irac1 and Irac2 (1.9~arcsec).  We show in Fig.~\ref{shiftres} the
differences between true and measured magnitudes as a function of
object size for the various filters. These shifts are most of the time
lower than 0.2 magnitude, except for the Irac1 and Irac2 bands. At the
median object size, Irac1 and Irac2 magnitude shifts are of the order
of 0.95, in perfect agreement with the previously quoted shifts
estimated with LePhare.

\subsection{Blended objects}

Objects which are nearby ($\sim 1$\arcsec) but separated in
ACS images may be blended in ground based or Spitzer images. Using
SExtractor in double image mode avoids the incorrect identification of
faint ACS detected objects with incorrect objects in the ground based
or Spitzer images.  We calculate the flux inside the Kron
aperture in the poorer spatial resolution images as determined from
the higher spatial resolution image (HST) at the exact place of the
faint object. This procedure limits the cross talk between the
fluxes. Furthermore, we flag these blends so we can determine if
including the objects or not changes the result we derive from our
\phz\ catalogs in a statistically significant manner. However, as we
show below, the quality of the \phzs\ for the blends is close to
those derived from the non-blended objects.

\subsection{Photo-$z$ accuracy with the spectroscopic sample}

\begin{figure}[!h]
  \begin{center}
    \includegraphics[angle=270,width=3.0in]{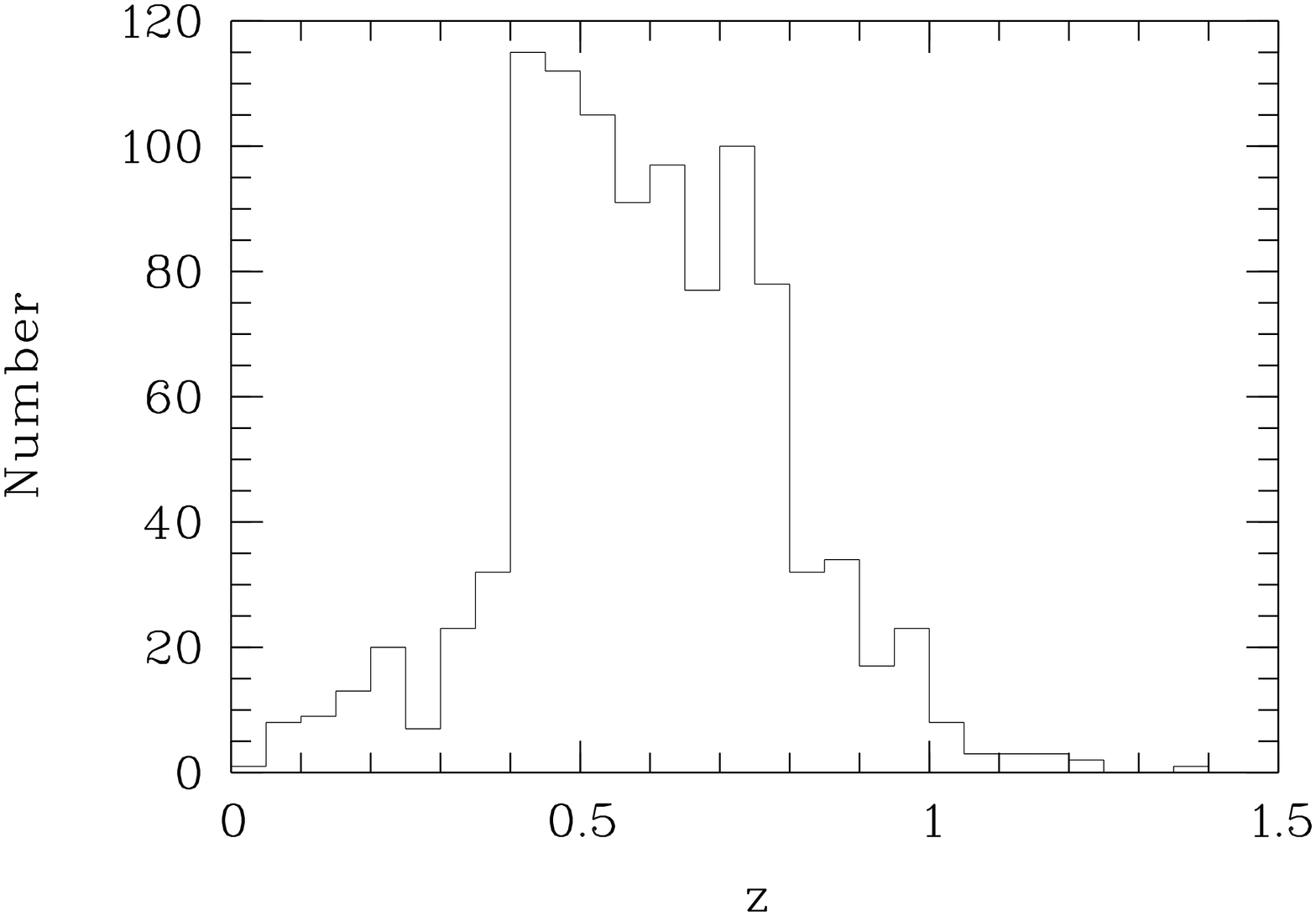}
    \includegraphics[angle=270,width=3.0in]{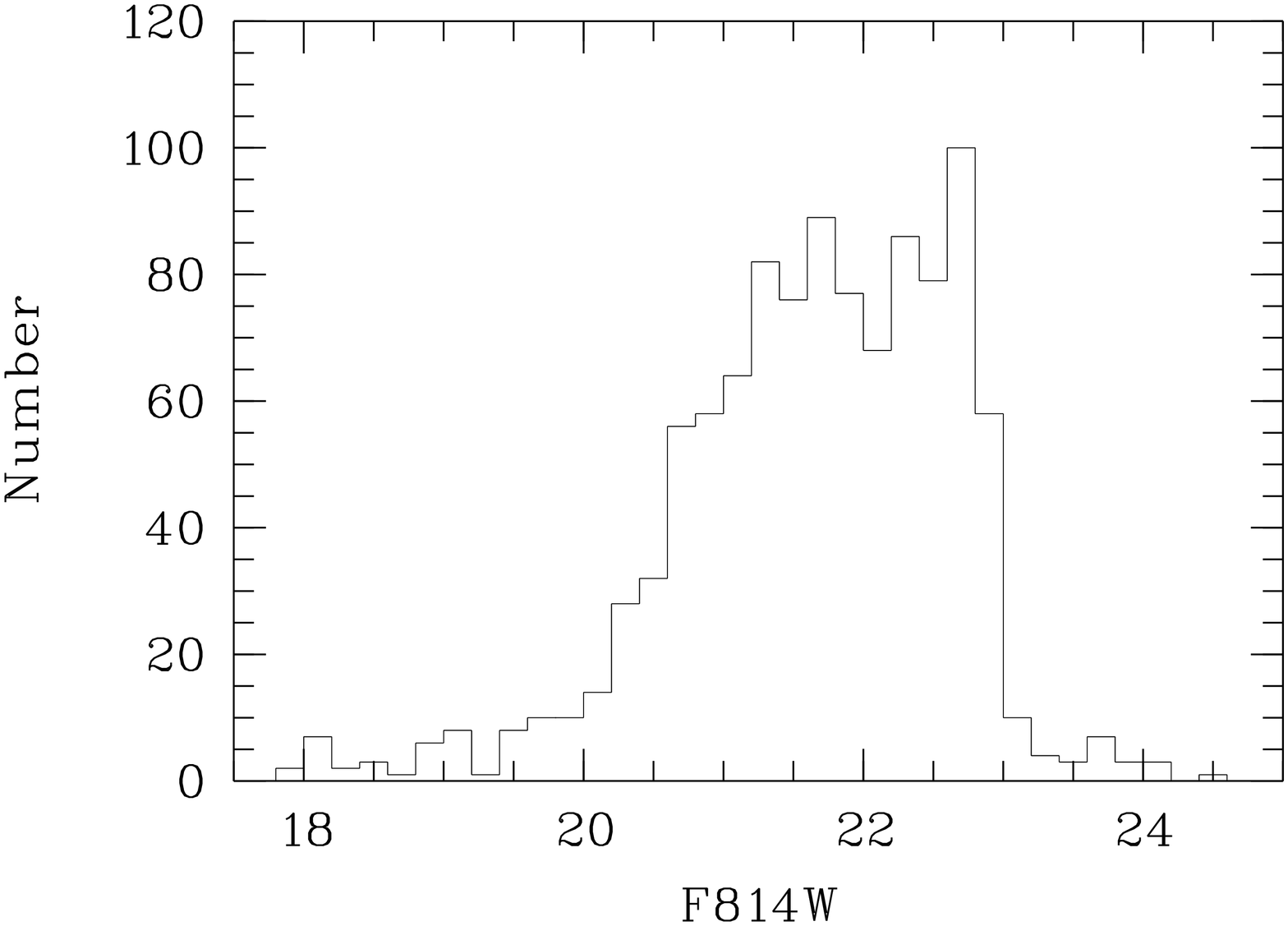}
    \caption{Redshift (top) and F814W magnitude (bottom) histograms of
      the spectroscopic sample.}
  \label{range}
  \end{center}
  \end{figure}

Given the results of Coupon et al. (2009), our data should be able to
provide good \phzs\ for magnitudes at least as faint as i$'$=24 (AB
magnitudes) and for redshifts lower than $\sim$1.5. 
We first estimate the quality of our \phzs\ by comparing them to
available spectroscopic redshifts. This gives us insight on the
\phz\ quality in the magnitude and redshift ranges covered by
the spectroscopic catalogs. We show in Fig.~\ref{range} the redshift
and F814W band magnitude histograms of the available spectroscopic
redshifts along the ten considered lines of sight. The \phz\ quality
assessment will be valid up to z$\sim$1.05 and for magnitudes brighter
than $\sim$23.5 and really strong  up to z$\sim$1 and for 
magnitudes brighter than $\sim$23. 

\begin{figure*}
  \centerline{
    \mbox{\includegraphics[angle=270,width=3.0in]{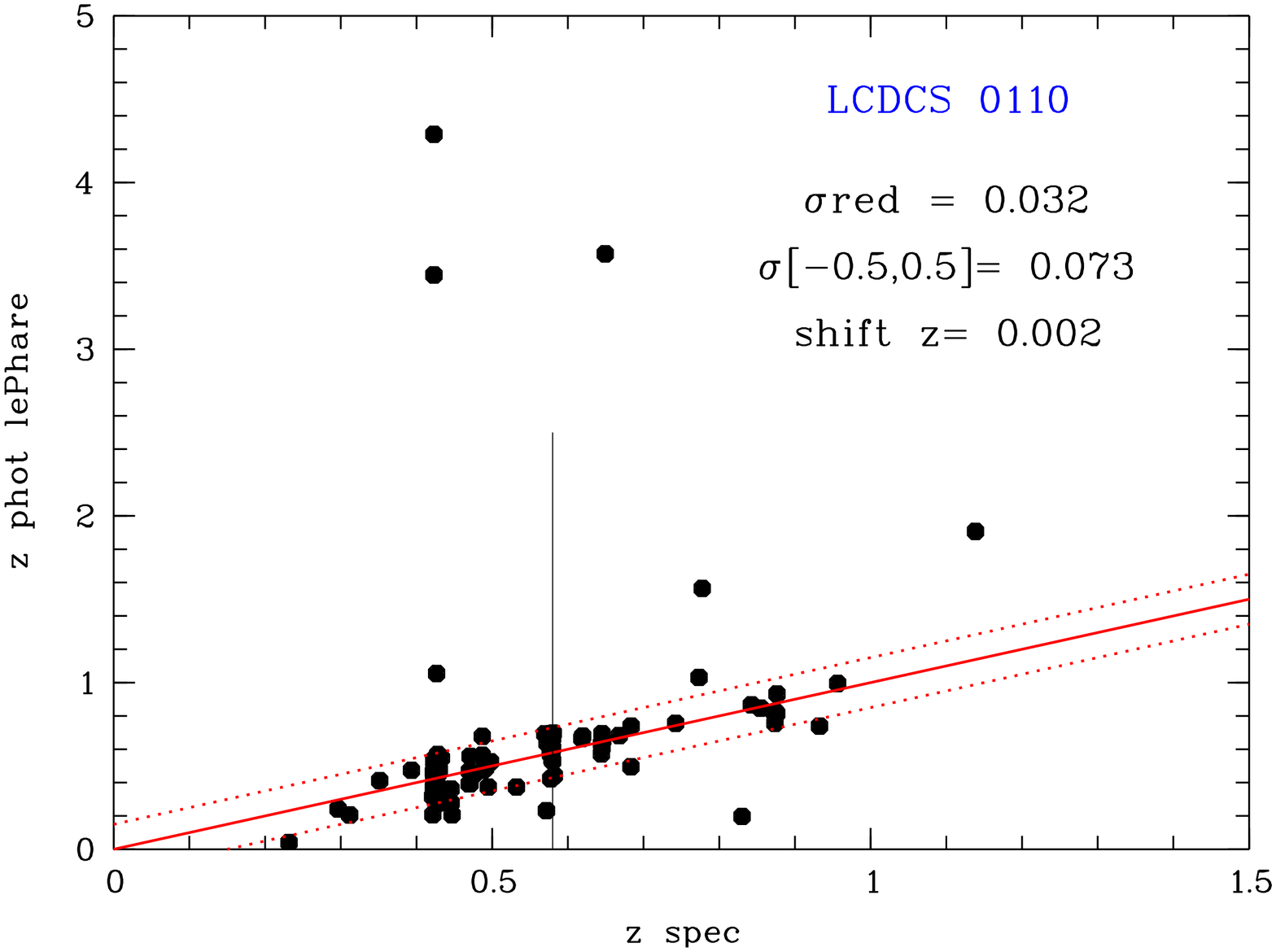}}
    \mbox{\includegraphics[angle=270,width=3.0in]{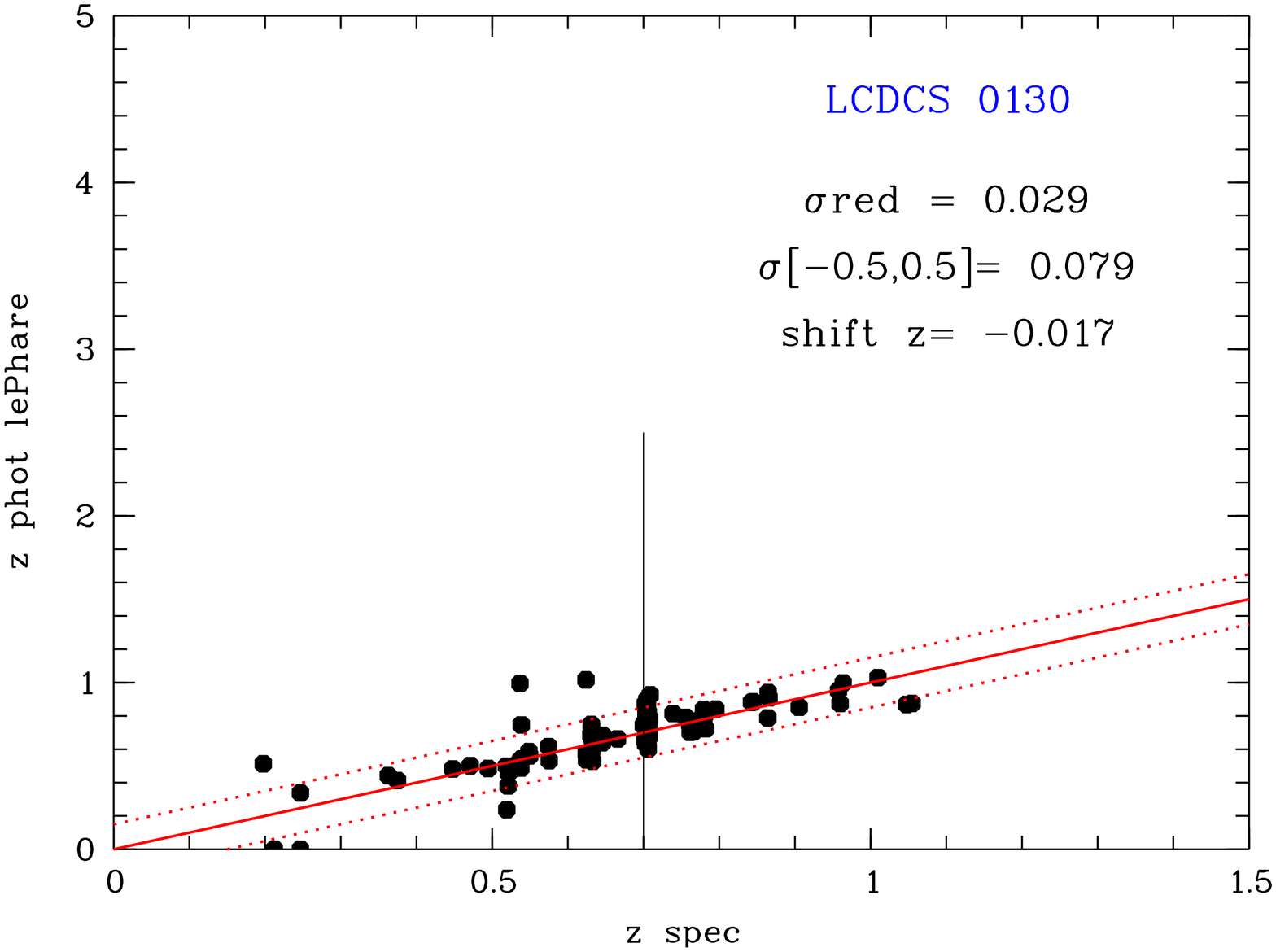}}
  }
  \centerline{
    \mbox{\includegraphics[angle=270,width=3.0in]{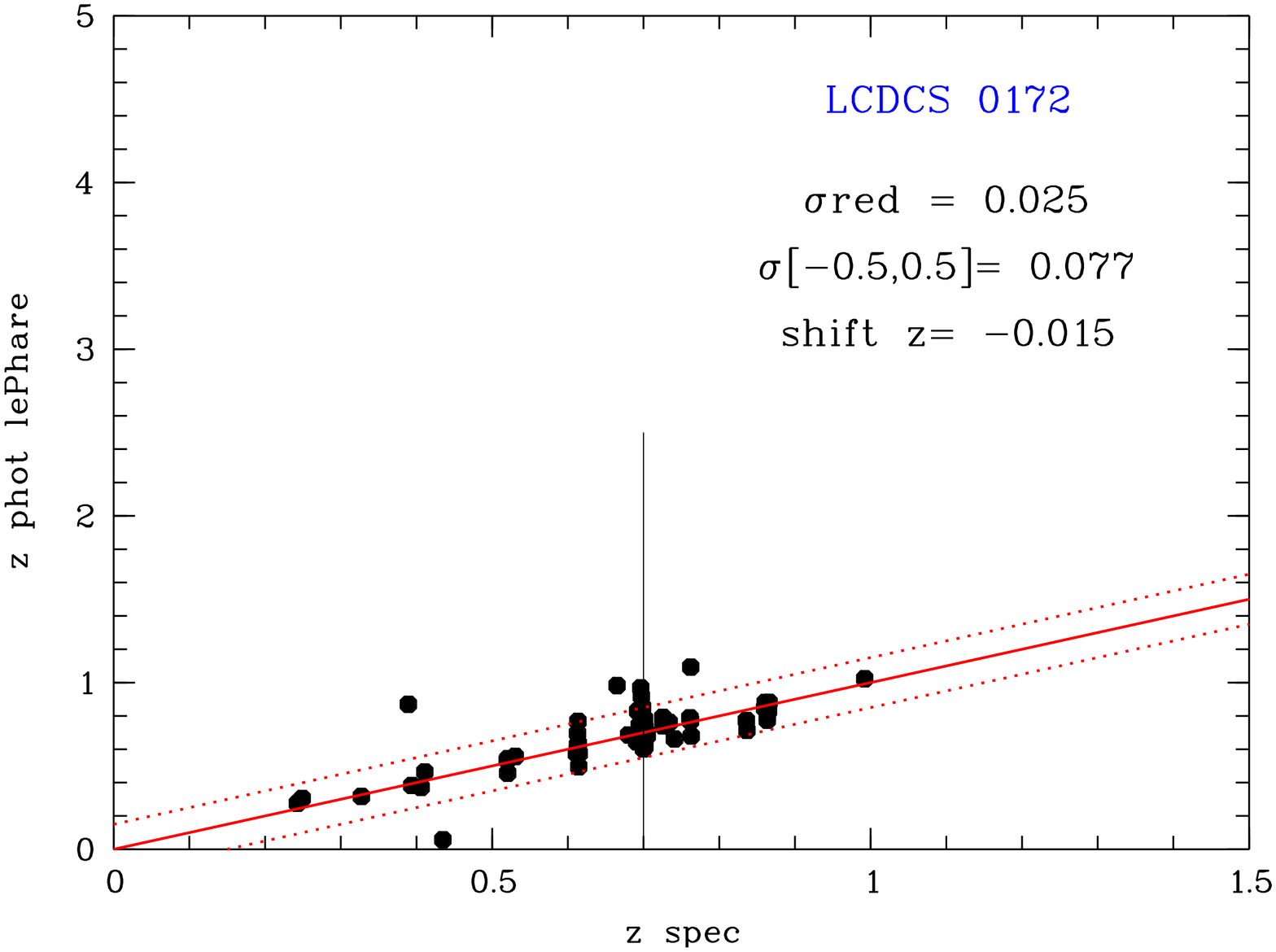}}
    \mbox{\includegraphics[angle=270,width=3.0in]{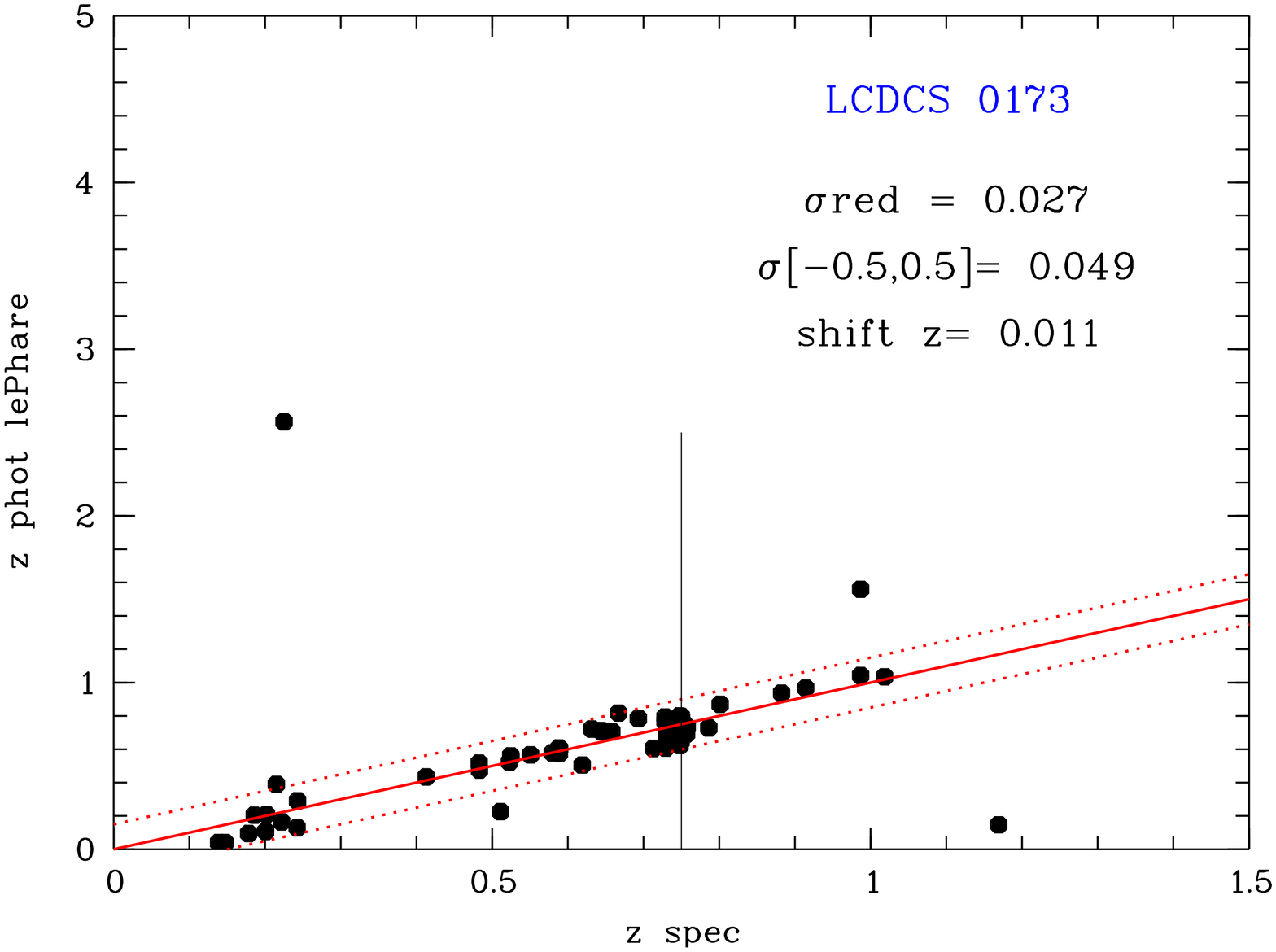}}
  }
  \caption{Spectroscopic versus photometric redshifts for 4
    clusters. We also give the dispersions around the mean relation
    (reduced value, classical value excluding galaxies for which the
    difference between spectroscopic and photometric redshifts is
    greater than 0.5), and the mean shift between \phzs\ and
    $specz$s. The solid inclined lines give the perfect relation
      while the dotted lines give the $\pm$0.15 relations. The
      vertical lines give the position of the cluster along the line
      of sight.}
  \label{zspeczphot}
  \end{figure*}

\begin{figure*}
  \centerline{
    \mbox{\includegraphics[angle=270,width=3.0in]{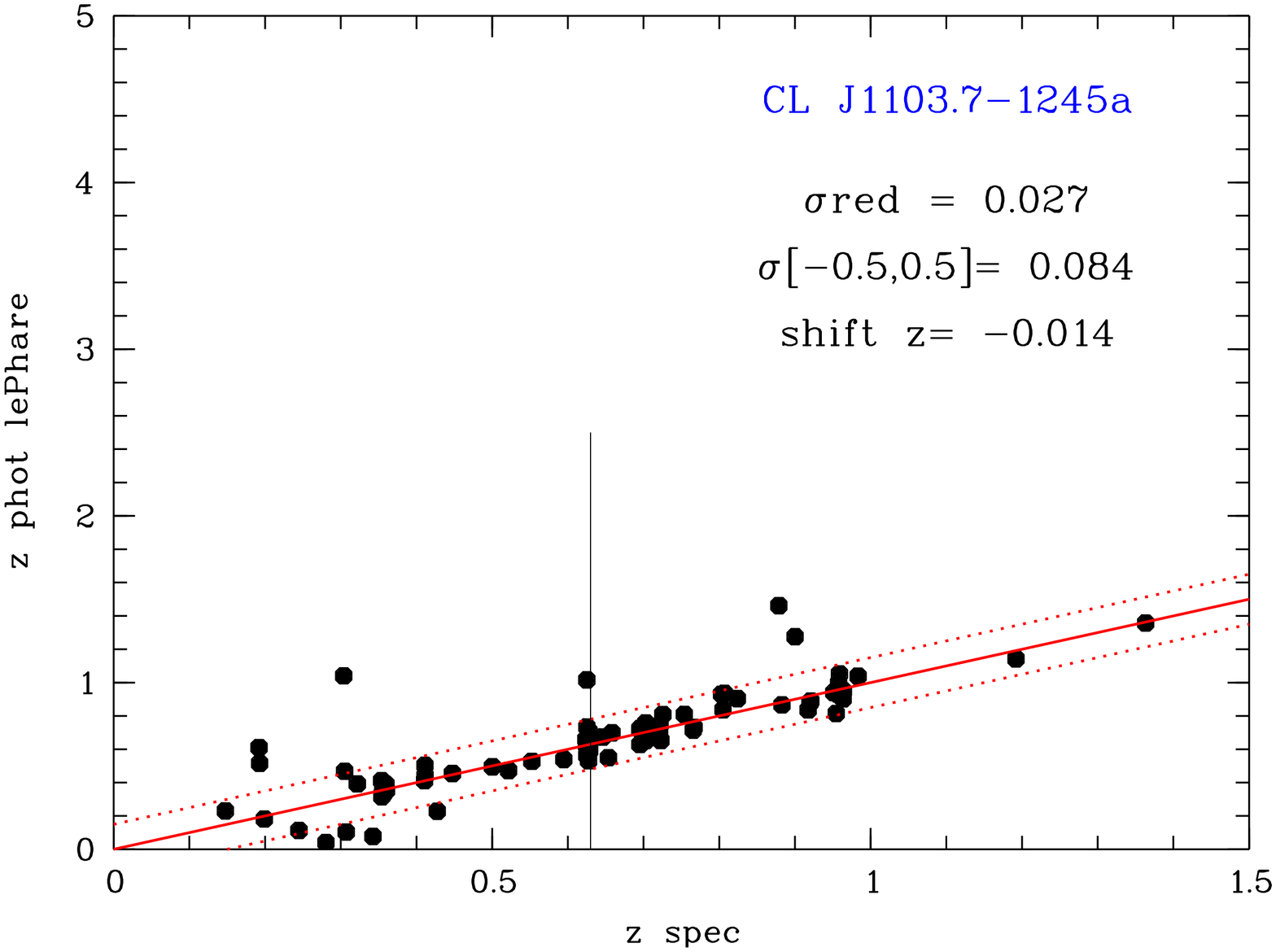}}
    \mbox{\includegraphics[angle=270,width=3.0in]{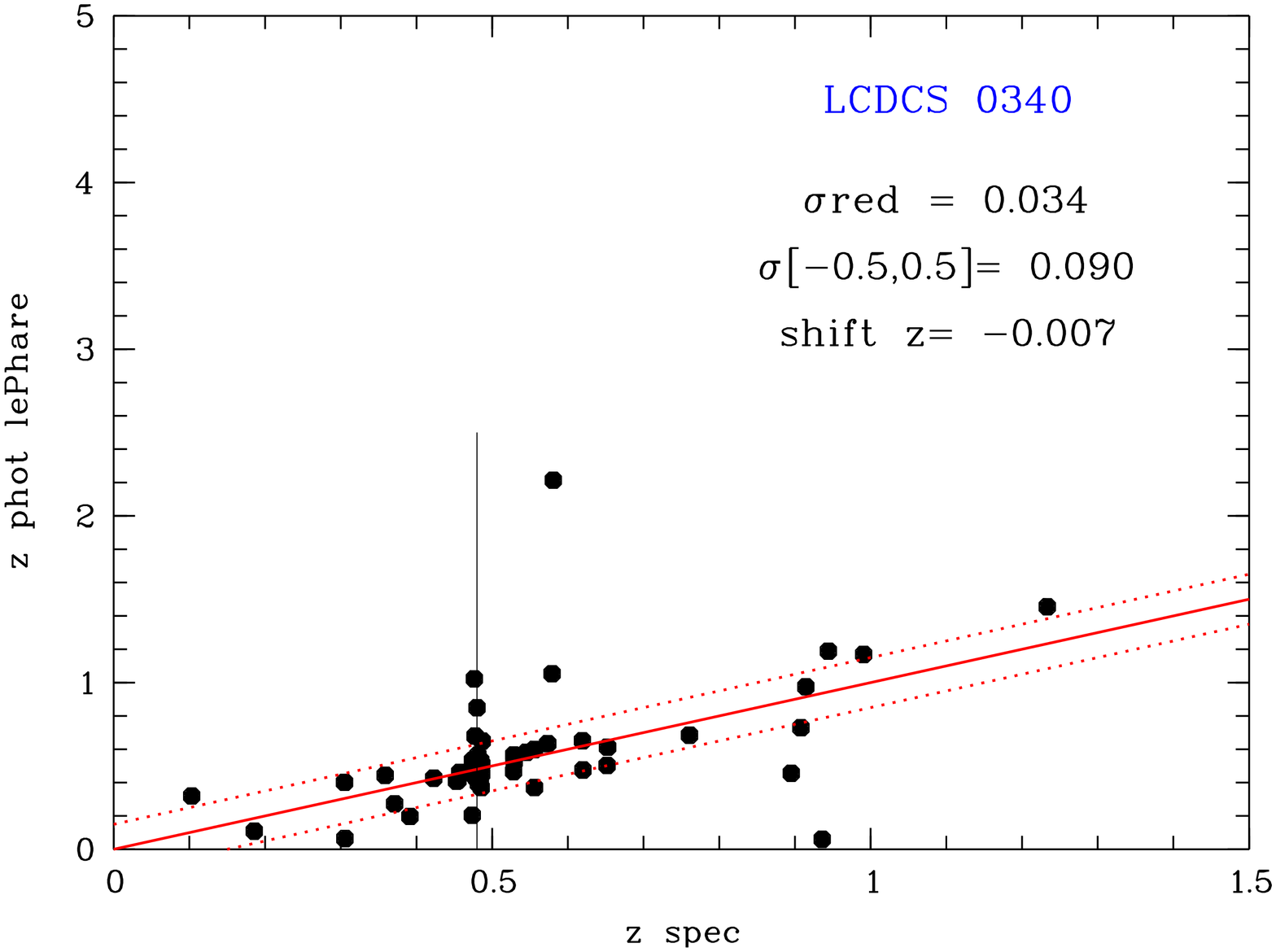}}
  }
  \centerline{
    \mbox{\includegraphics[angle=270,width=3.0in]{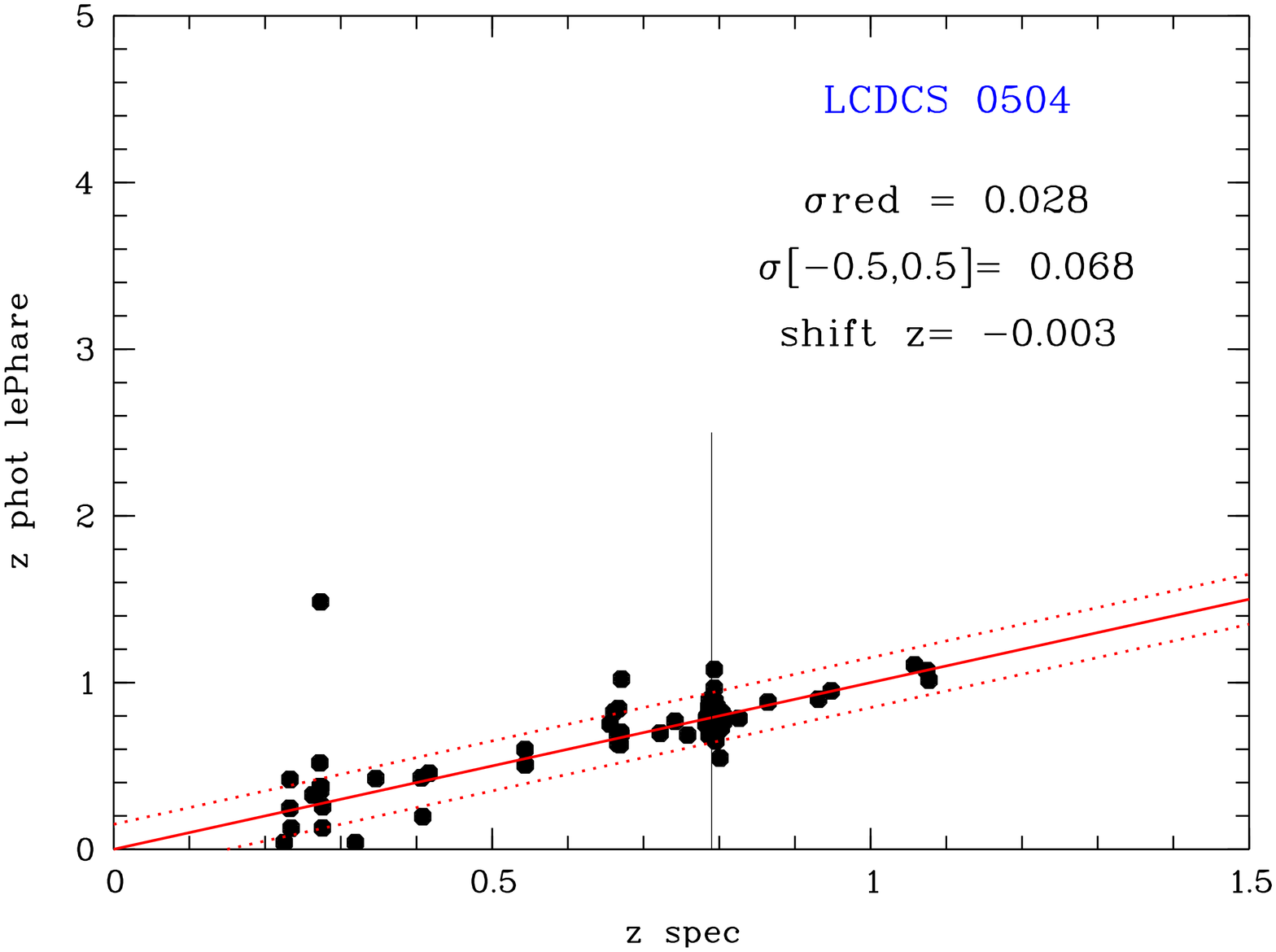}}
    \mbox{\includegraphics[angle=270,width=3.0in]{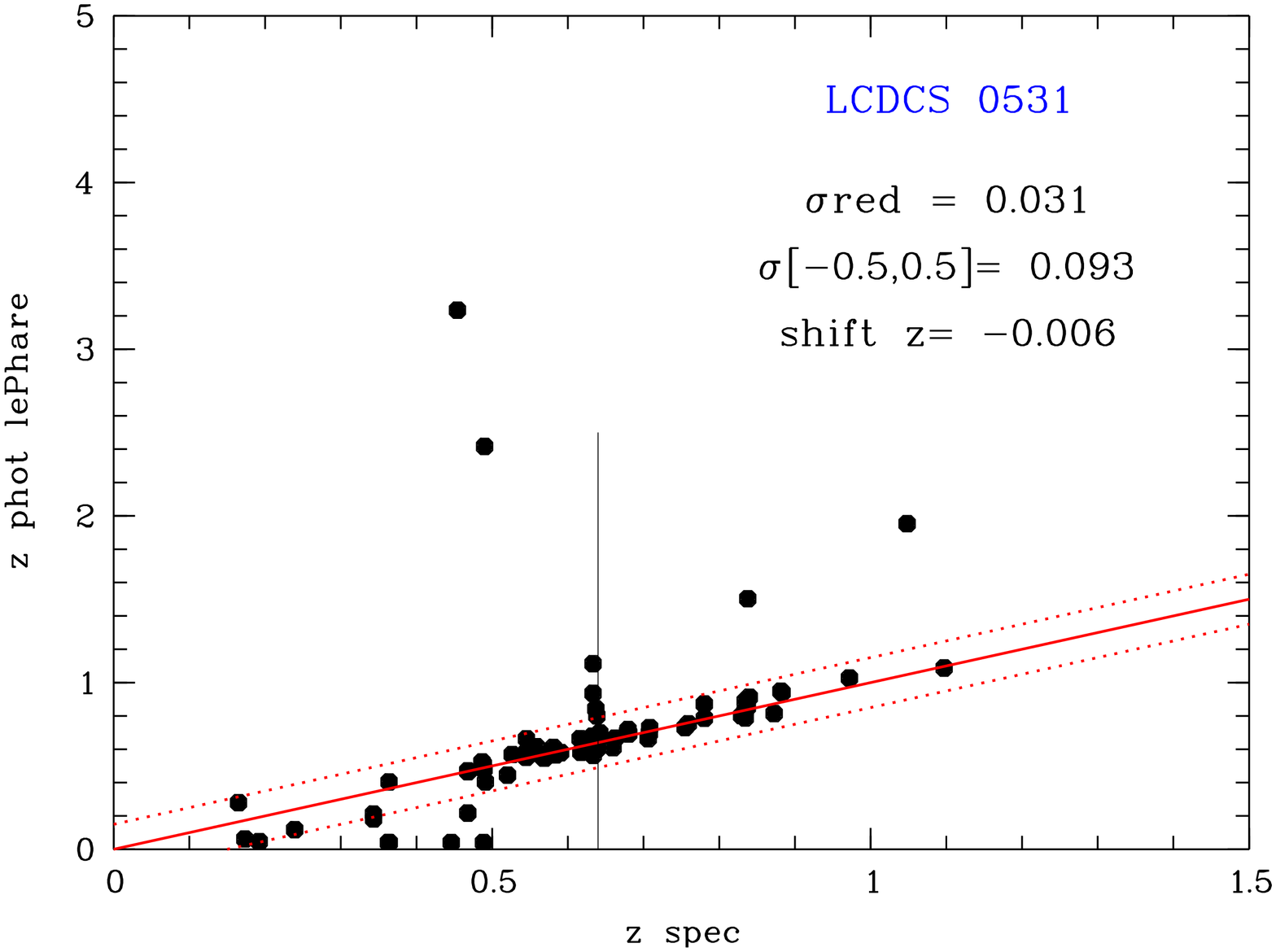}}
  }
  \centerline{
    \mbox{\includegraphics[angle=270,width=3.0in]{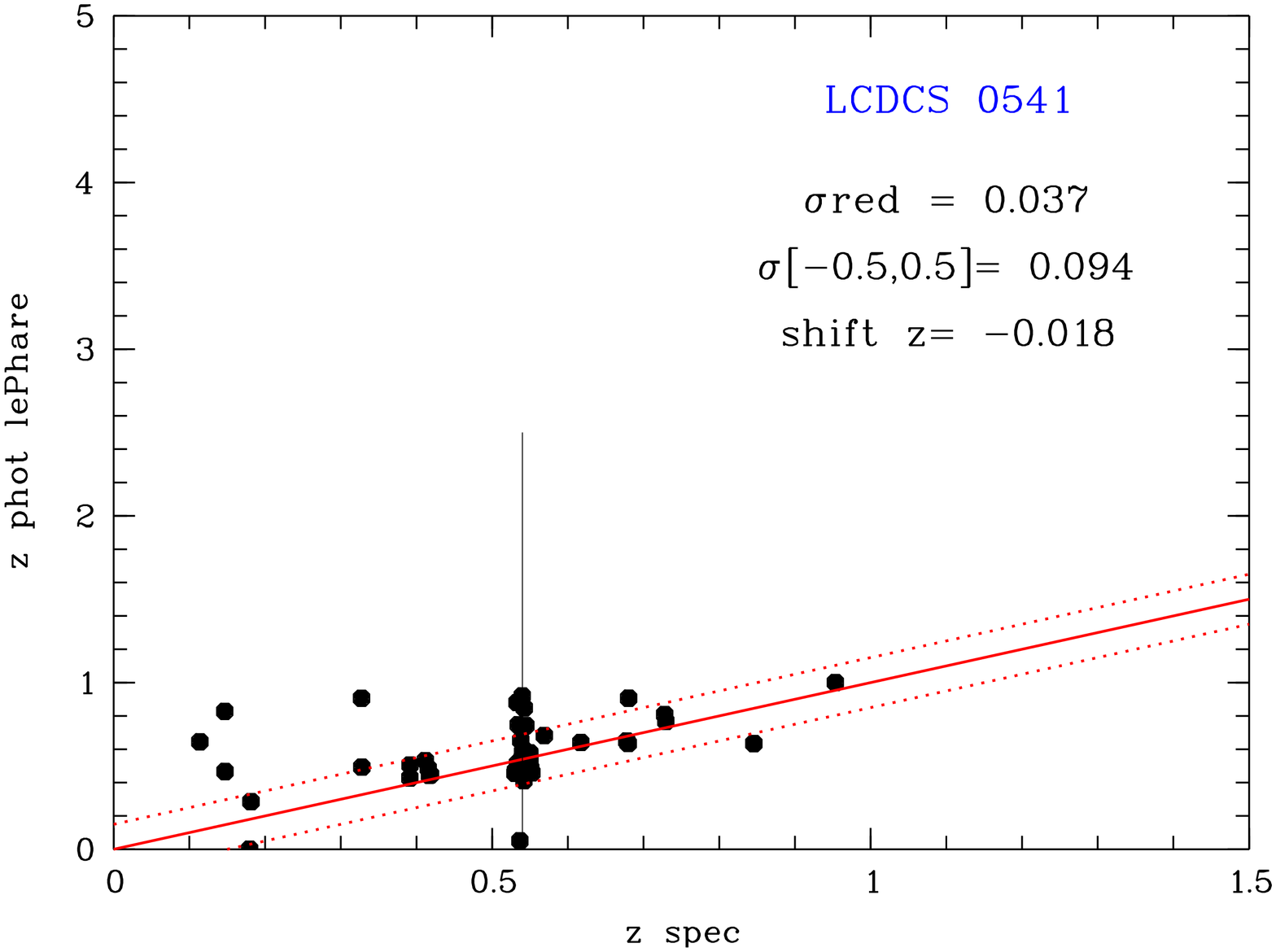}}
    \mbox{\includegraphics[angle=270,width=3.0in]{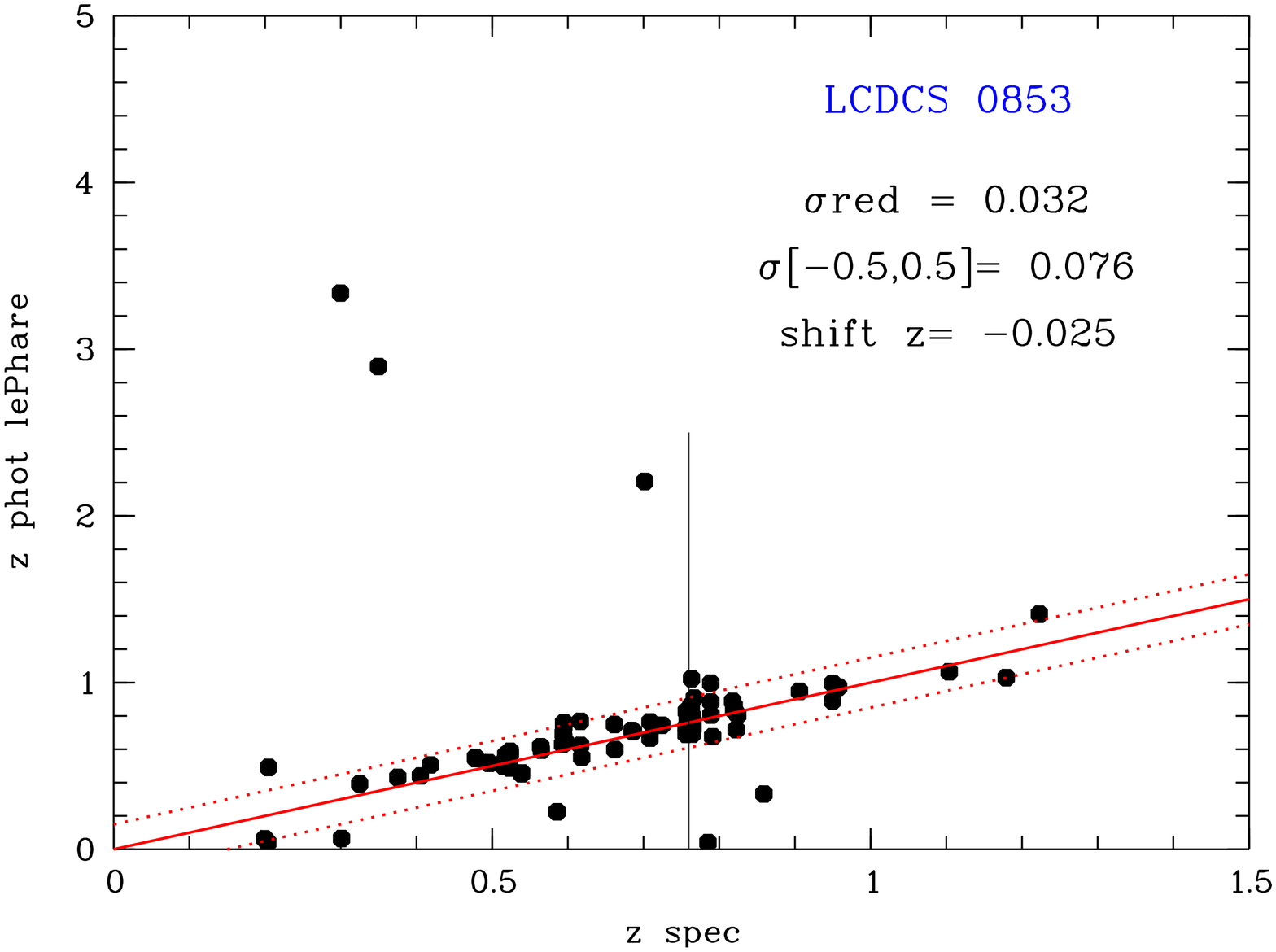}}
  }
  \caption{Same as Fig.~\ref{zspeczphot} for the 6 other clusters.}
  \label{zspeczphot2}
  \end{figure*}

  We then generated the \phz\ versus spectroscopic redshift ($specz$
  hereafter) plots shown in Figs.~\ref{zspeczphot} and
  ~\ref{zspeczphot2}. These plots give the dispersions around the mean
  relation (both the NMAD reduced sigma of Ilbert et al. (2006) 
    \footnote{1.48 $\times $ median( $\vert \Delta z\vert/(1+z))$} and the
  regular sigma value: second moment of the value distribution), and the
  mean shift between \phzs\ and $specz$s.  We clearly see
  that on average, \phzs\ and $specz$s are in good agreement with a
  reduced $\sigma$ of the order of 0.05 around the mean relation
  (regular sigma of the order of 0.09). The percentage of catastrophic
  errors (objects with a difference 
  between true redshift and \phz\ of more than 0.2$\times$(1+$specz$)) is not 
  negligible, but remains lower than 5$\%$. It is also tempting to see an increased
  uncertainty in the \phz\ estimates when considering cluster galaxies (except for 
  LCDCS 0173). We will however quantify this possible effect later in the paper.
  We also remark that catastrophic errors mainly occur towards the high photometric 
  redshifts (at z$\geq$1.5) and we will discuss the possible consequences on our survey 
  in the final section.

\begin{figure}[!h]
  \begin{center}
    \includegraphics[angle=270,width=3.00in]{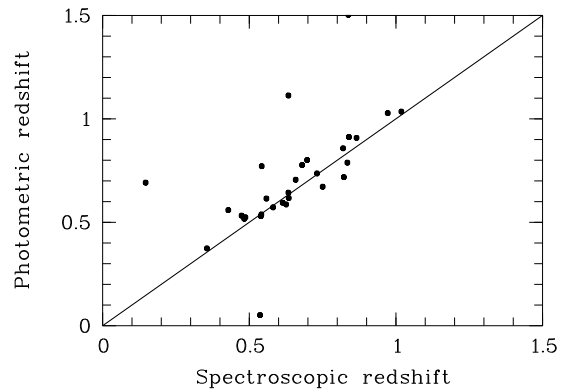}
    \caption{Photometric versus spectroscopic redshifts for the
      blended objects in the spectroscopic sample.}
  \label{blendspec}
  \end{center}
\end{figure}

  The question is now to know if \phzs\ of blended objects are also
  acceptable. We therefore flagged all such objects by selecting
  galaxies with a close neighbor (at less than 1.5~arcsec, given the
  ground based seeings) and less than 0.5 magnitude fainter than the
  primary object (enough to potentially bias the magnitude
  estimate). Such objects are potentially polluted by a comparable or
  brighter object (less than 0.5 magnitude fainter, or brighter). We
  then generated Fig.~\ref{blendspec} where all such objects from the
  ten considered clusters are shown. The reduced $\sigma$ is 0.08. The
  percentage of catastrophic errors is $\sim$10\%, higher than for the
  whole sample of galaxies (blended or not). However even if these two
  values (the reduced $\sigma$ and the percentage of catastrophic
  errors) are higher than for the whole sample, they remain acceptable
  for our purposes and we conclude that blending is not a redhibitory
  problem for z$\leq$1.05 and F814W$\leq$23.5.

\subsection{Photo-$z$ accuracy and environmental dependence}

Although the first goal for the FADA/DAFT project is to determine the
\phzs\ of background field galaxies, it is also of interest to
determine the achieved \phz\ accuracy and the fraction of catastrophic 
errors as a function of various cluster/galaxy internal
parameters. In clusters, Adami et al. (2010) have already shown with
the X-ray selected XMM-LSS sample that late type galaxies tend to
exhibit poorer \phz\ precision than early type galaxies.  Moreover,
the XMM-LSS clusters are not very massive structures and environmental
effects are perhaps not as strong as in the presently considered
clusters.

With the sample presently in hand, we determined the effect of galaxy
type and magnitude on the \phz\ accuracy in the redshift range for
which spectroscopic data exist. Similarly, we can investigate the
possible effects of the environment (cluster versus field regions).

\begin{figure}
  \begin{center}
    \includegraphics[angle=270,width=3.00in]{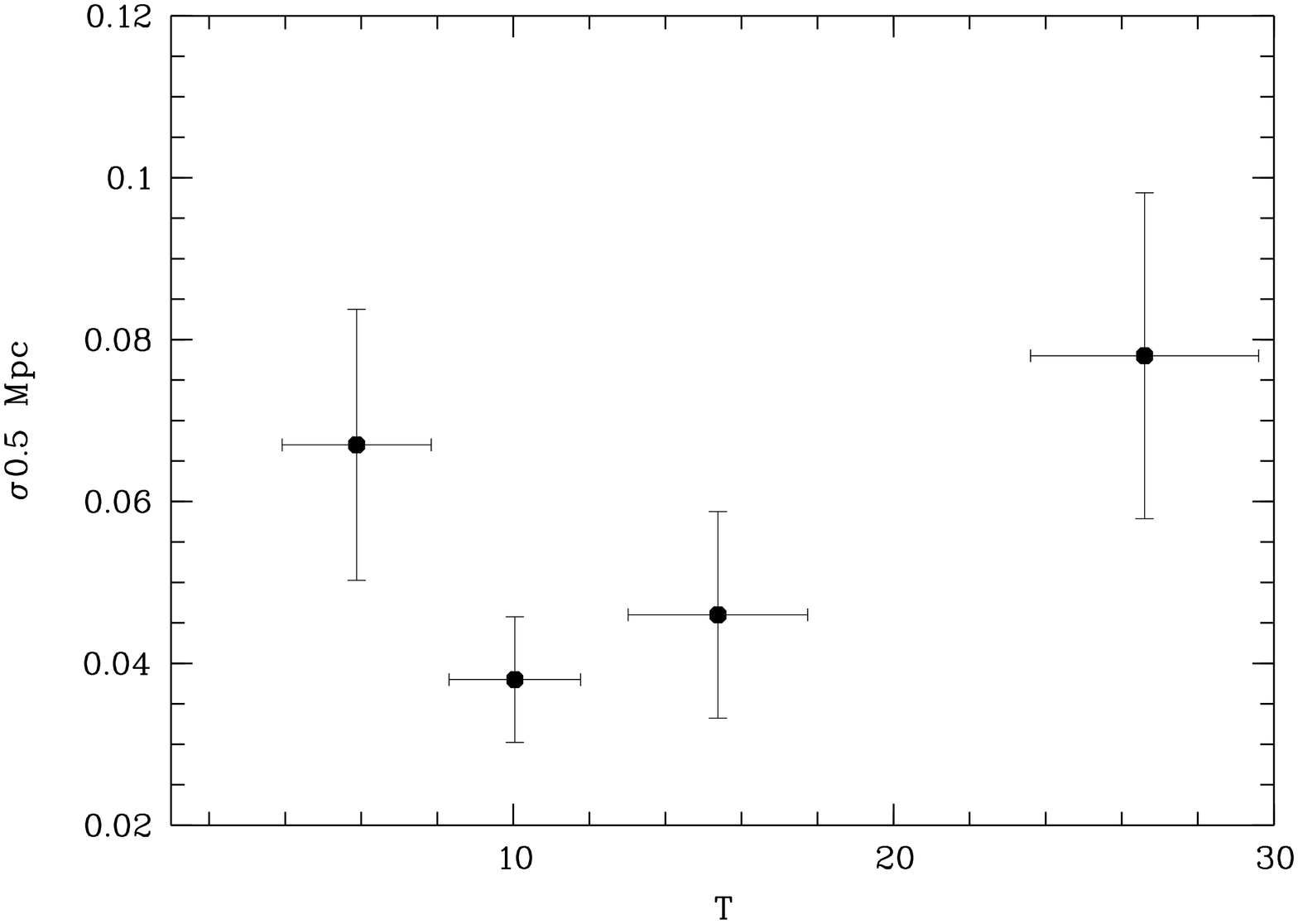}
    \includegraphics[angle=270,width=3.00in]{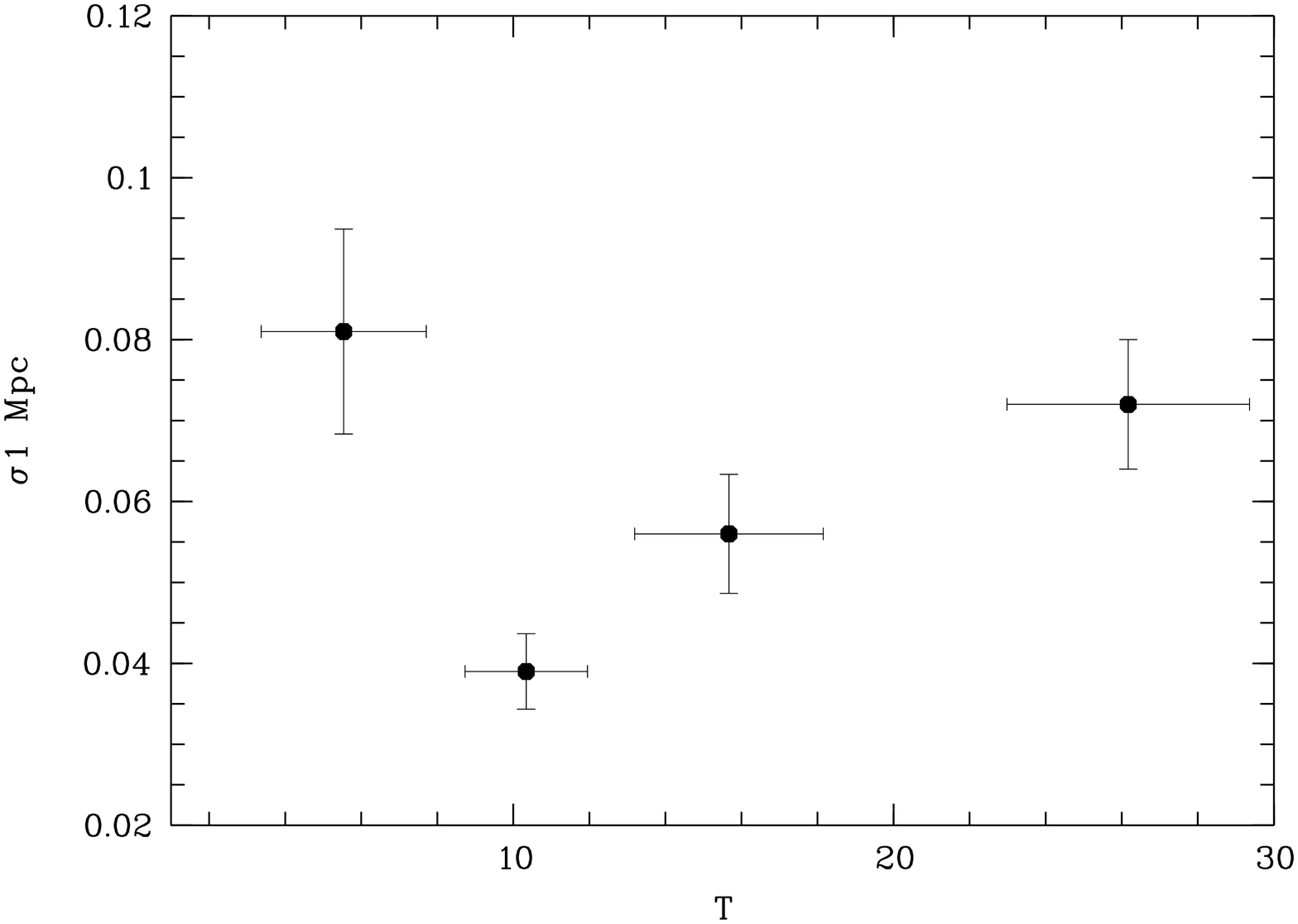}
    \includegraphics[angle=270,width=3.00in]{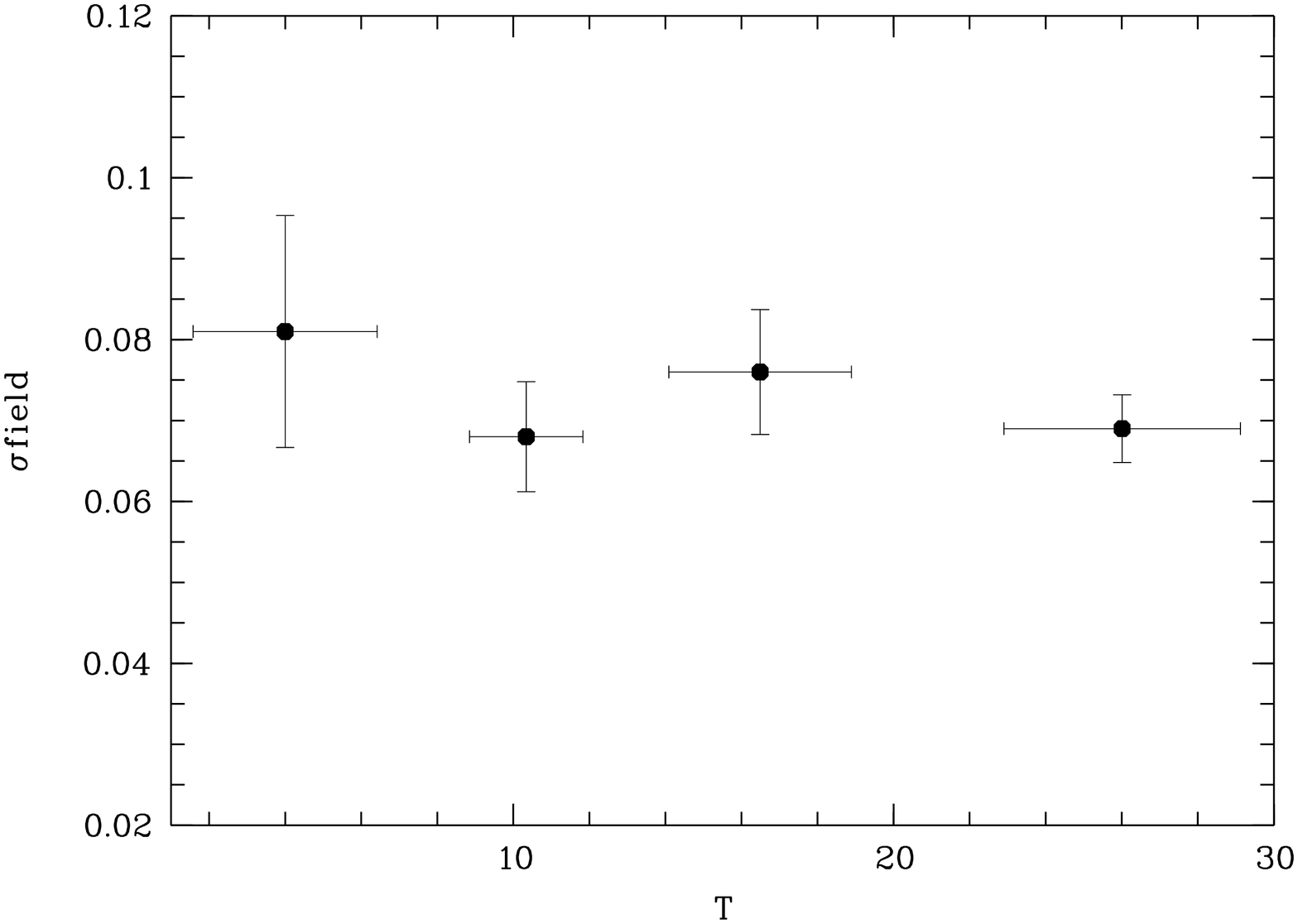}
    \caption{Reduced $\sigma$ of \phzs\ versus galaxy photometric type
      $T$. From top to bottom: cluster
      galaxies inside a 500 kpc radius, inside a 1~Mpc radius, and
      field galaxies. Error bars for the types are simply the second order 
      momentum of the galaxy type distribution in the selected type bins 
      ([1;7], [8;12], [13;19], [20;31]). Error bars for the reduced $\sigma$
    are Poissonian error bars and are therefore directly proportional to
    the inverse of the number of galaxies inside the considered bin.}
  \label{precT}
  \end{center}
  \end{figure}

\begin{figure}
  \begin{center}
    \includegraphics[angle=270,width=3.00in]{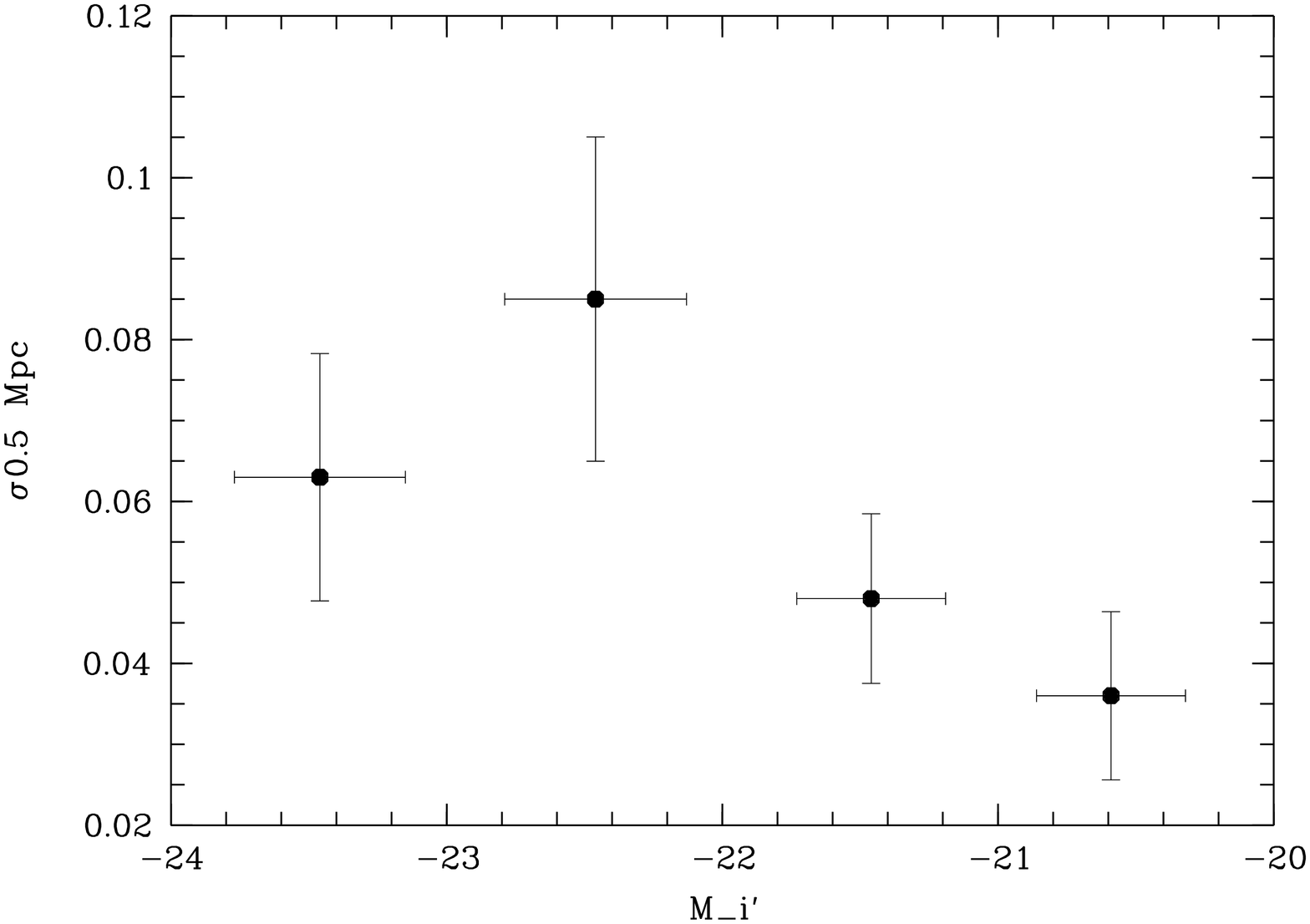}
    \includegraphics[angle=270,width=3.00in]{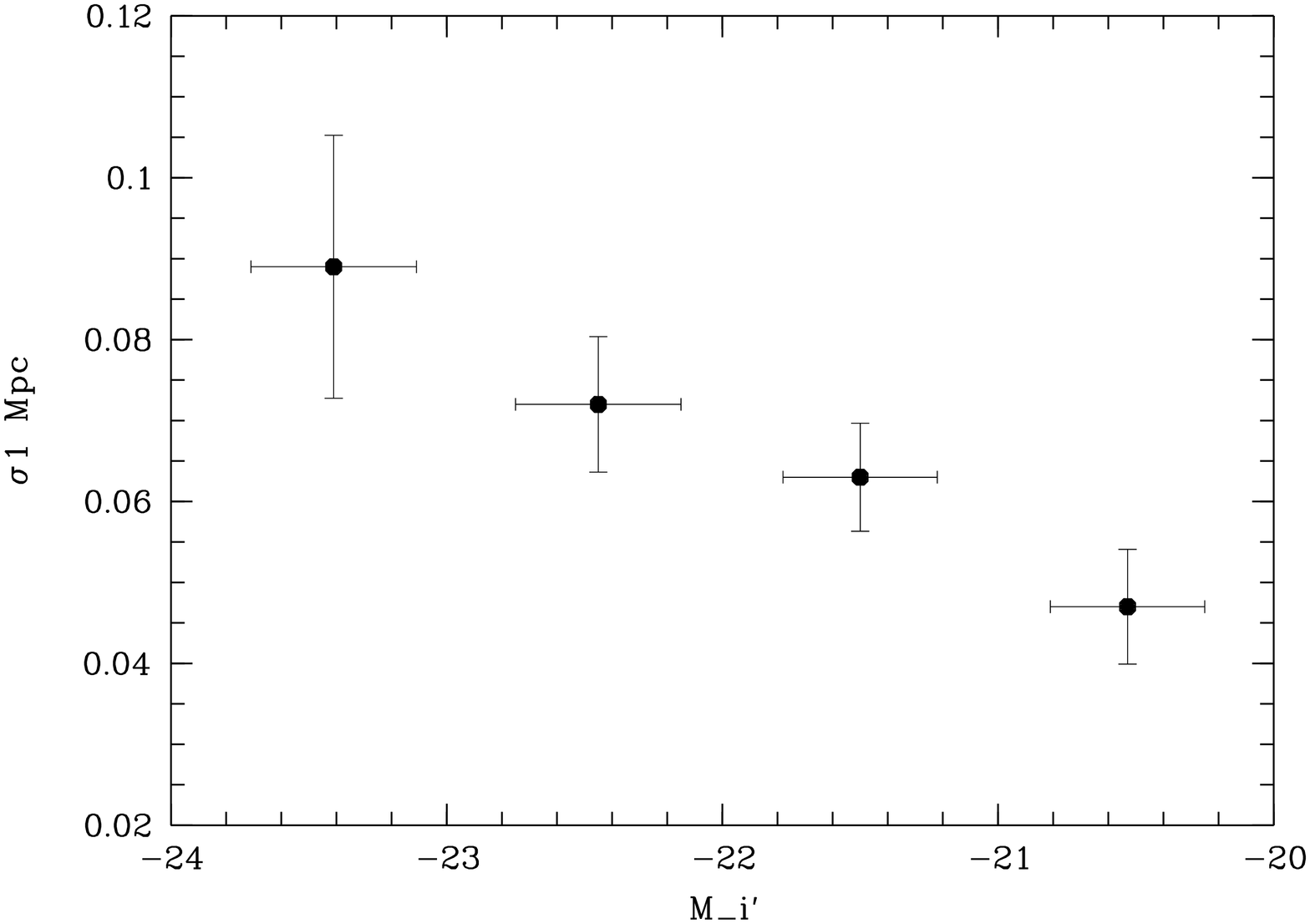}
    \includegraphics[angle=270,width=3.00in]{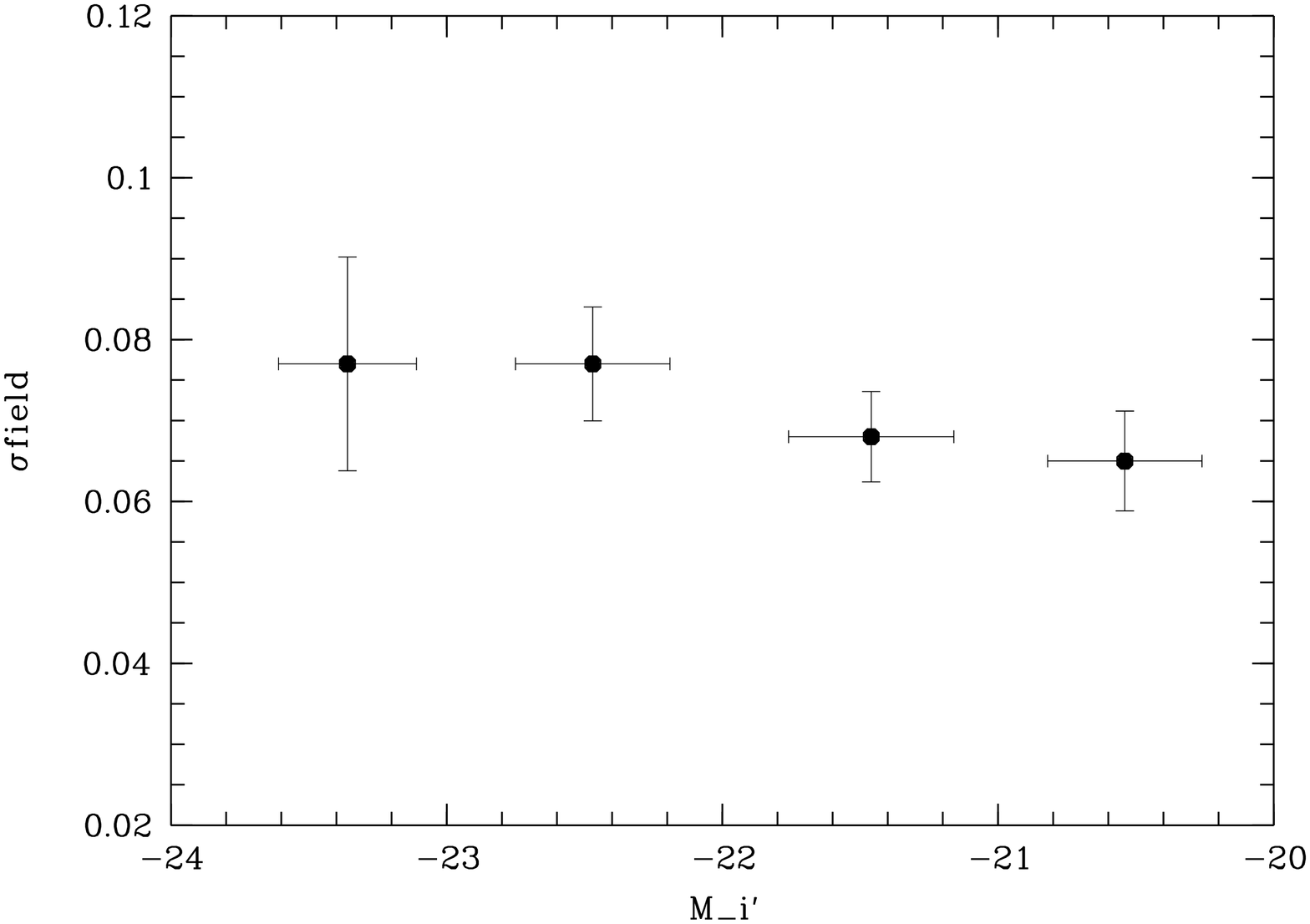}
    \caption{Reduced $\sigma$ of \phzs\ versus galaxy i' absolute
      magnitude. From top to bottom: cluster
      galaxies inside a 500~kpc radius, inside a 1~Mpc radius, and
      field galaxies. Error bars for the absolute magnitudes are simply the 
      second order momentum of the galaxy magnitude distribution in the 
      selected magnitude bins ($[-24;-23], [-23;-22], [-22;-21], [-21;-20]$). 
      Error bars for the reduced $\sigma$
    are Poissonian error bars and are therefore directly proportional to
    the inverse of the number of galaxies inside the considered bin.}
  \label{precM}
  \end{center}
  \end{figure}

We chose to merge the 10 clusters in a single photometric versus
spectroscopic redshift catalog. The following results will therefore
apply for the considered cluster redshift range (z~[0.4;0.9]). We show
in Figs.~\ref{precT} and ~\ref{precM} the variation of the reduced
$\sigma$ of the \phzs\ as a function of photometric type and of
absolute magnitude for cluster galaxies within a projected
cluster-centric radius of 0.5~Mpc and 1~Mpc, and field
galaxies. Galaxies were selected as members of the 1~Mpc radius region
if their redshift differed by less than 3 times the velocity
dispersion of Clowe et al. (2006) compared to the mean cluster
redshift.  For the 0.5~Mpc region, the limit was set to less than one
time the velocity dispersion of Clowe et al. (2006).

  These figures show that we have a worse \phz\ accuracy for the
  brightest cluster galaxies (in absolute magnitude).  Differences
  between best and worst values represent most of the time a factor of
  two, in good agreement with the results of Adami et al. (2010).  The
  tendency is clearly different in the field where the variation is
  not significant. Considering now the photometric type $T$, we see in
  Fig.~\ref{precT} the worse \phz\ accuracy for the latest and
  earliest type objects in cluster regions while again field galaxies
  do not show any significant tendency.

  Considering now catastrophic error percentages, in clusters such errors 
  occur for both bright and late
  type galaxies. Table~\ref{tbladd2} gives the spectral type and 
  magnitude intervals for which we have non null 
  catastrophic error percentages in clusters. In the field, these
  percentages are non null whatever the galaxy type or magnitude and
  there is no clear trend to have preferably high or low catastrophic
  error percentages for specific galaxy types or magnitudes.

\begin{table*}
\caption{Spectral type and magnitude intervals for a given cluster area for which
the percentage of catastrophic errors (4th column) is non null.}
\label{tbladd2}
\begin{center}
%\begin{tabular}{|c|c|c|c|c|c|} 
\begin{tabular}{ccccccc} 
\hline
Spectral type & Magnitude & Radius & Percentage \\ \hline
  $19\leq T \leq 31$ & M$_{i'}=[-23,-22]$ & 1Mpc & 11$\pm$11$\%$ \\ 
  $19\leq T \leq 31$ & M$_{i'}=[-25,-23]$ & 1Mpc & 28$\pm$19$\%$ \\ 
  $19\leq T \leq 31$ & M$_{i'}=[-22,-21]$ & 0.5Mpc & 5$\pm$4$\%$   \\ 
  $19\leq T \leq 31$ & M$_{i'}=[-25,-22]$ & 0.5Mpc & 19$\pm$7$\%$  \\ 
  $12\leq$$T$$\leq19$ & M$_{i'}=[-23,-22]$ & 0.5Mpc & 5$\pm$5$\%$   \\ 
\hline
\end{tabular}
\end{center}
\end{table*}

\begin{figure}[!h]
  \begin{center}
  \includegraphics[angle=270,width=3.00in]{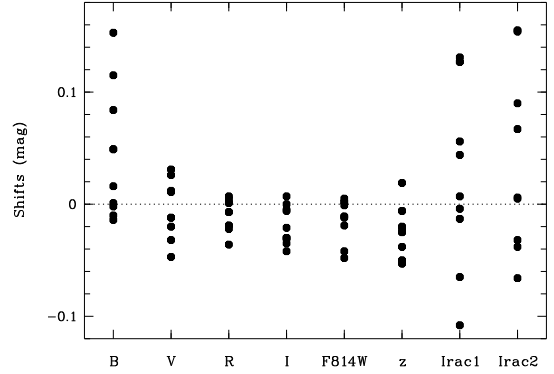}
  \caption{Difference between LePhare estimated magnitude shifts (the
    value when considering only cluster galaxy redshifts minus the
    value when considering all available galaxy redshifts) for the
    various photometric bands considered. A positive value means that
    magnitudes have to be more weakened when considering cluster
    galaxy redshifts than when considering all available redshifts
    (from cluster or field galaxies).}
  \label{shiftdiff}
  \end{center}
\end{figure}

\begin{figure}[!h]
  \begin{center}
    \includegraphics[angle=270,width=3.00in]{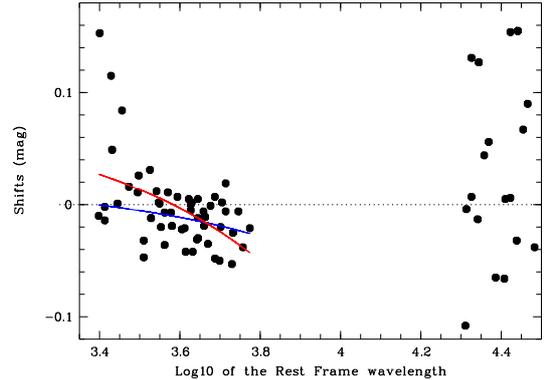}
    \caption{Same as Fig.~\ref{shiftdiff} at the rest frame wavelength. The red inclined 
line is the regression line for the visible bands. The blue inclined line is the regression 
line for the visible bands when considering only the shifts smaller than 0.04.}
  \label{shiftdiffz}
  \end{center}
\end{figure}

  How can these tendencies be understood? It is tempting to say that
  photometric redshift SEDs are environment dependent. This would not
  be surprising as the commonly used SEDs are adapted to low density
  environments and resulting \phz\ accuracy could be degraded when
  considering cluster galaxies. The spectral types showing in clusters
  the most atypical evolution compared to field objects are early type
  galaxies (cluster dominant galaxies) and very late type galaxies
  (galaxies with short bursts of star formation induced by the
  intracluster influence). These are exactly the ones showing the
  worst \phz\ accuracies in the previous tests. Moreover, during the
  process of the zero point shift estimates, we recall that we are
  comparing the photometric and spectroscopic redshifts used for
  training sets.  These training sets are dominated by field galaxies
  for most of the presently considered lines of sight ($\sim$60$\%$ of
  the available redshifts are field objects), so photometric redshifts
  for cluster galaxies may well not be optimally computed. In order to
  test these possibilities, we computed the shifts to apply to our
  photometry when including in spectroscopic training sets all the
  available redshifts or only those belonging to clusters.
  Fig.~\ref{shiftdiff} shows the difference between these shifts as a
  function of the considered photometric band disregarding the cluster
  redshift.  Fig.~\ref{shiftdiffz} shows the same shifts but at the
  rest frame wavelength (only for optical bands). This last figure is
  sensitive to the general SED shape. We clearly see that most of the
  optical magnitudes need to be brighter to reach the best SED when
  using cluster training sets, except for the B band which needs to be
  fainter. From Fig.~\ref{shiftdiffz}, we can also say that SEDs for
  cluster galaxies would need to be (sligthly) fainter at red
  wavelengths and brighter by $\sim$0.1 magnitude at blue
  wavelengths. The effect remains modest (less than 0.1 magnitude most
  of the time) but it is nearly systematic over the 10 considered
  lines of sight. The magnitude shifts computed with the global
  training sets are therefore not perfectly adapted to cluster
  galaxies, and used SEDs are also not very well adapted to cluster
  galaxies. This probably explains part of the \phz\ accuracy
  dependence on the environment. These results therefore confirm the
  need for high density environment SEDs when high precision \phzs\
  are required for cluster studies. The main improvement should
  essentially come from new SEDs for bright and very early or very
  late type cluster galaxies at various redshifts.

\subsection{Photo-$z$ quality checks: beyond the spectroscopic sample limits}

We now ask the question of the \phz\ quality beyond z=1.05 as well as
for objects fainter than F814W=23.5. These ranges cannot be tested
with the spectroscopic data in hand, so we chose the same approach as
in Ilbert et al. (2009). This paper shows that the percentage of
galaxies with an individual \phz\ error (estimated by LePhare) larger
than a given value is an indicator of the catastrophic error
percentages while the mean value of the individual \phz\ errors (for a
given object subsample) is a good approximation of the dispersion of
the $specz$/\phz\ relation (for the same subsample).

Before directly applying this approach, we first need to test it on
our data. For this, we select galaxies with a spectroscopic redshift
and we compute how many times the spectroscopic value falls inside the
1$\sigma$ interval given by the \phzs.  The result is
67$\pm$8\%, in good agreement with the expected 68\% for the 1$\sigma$
interval. We can therefore quantify the global \phz\ accuracy based on
the 1$\sigma$ error bars of individual \phzs.

\begin{figure}
  \centerline{
    \mbox{\includegraphics[angle=270,width=1.8in]{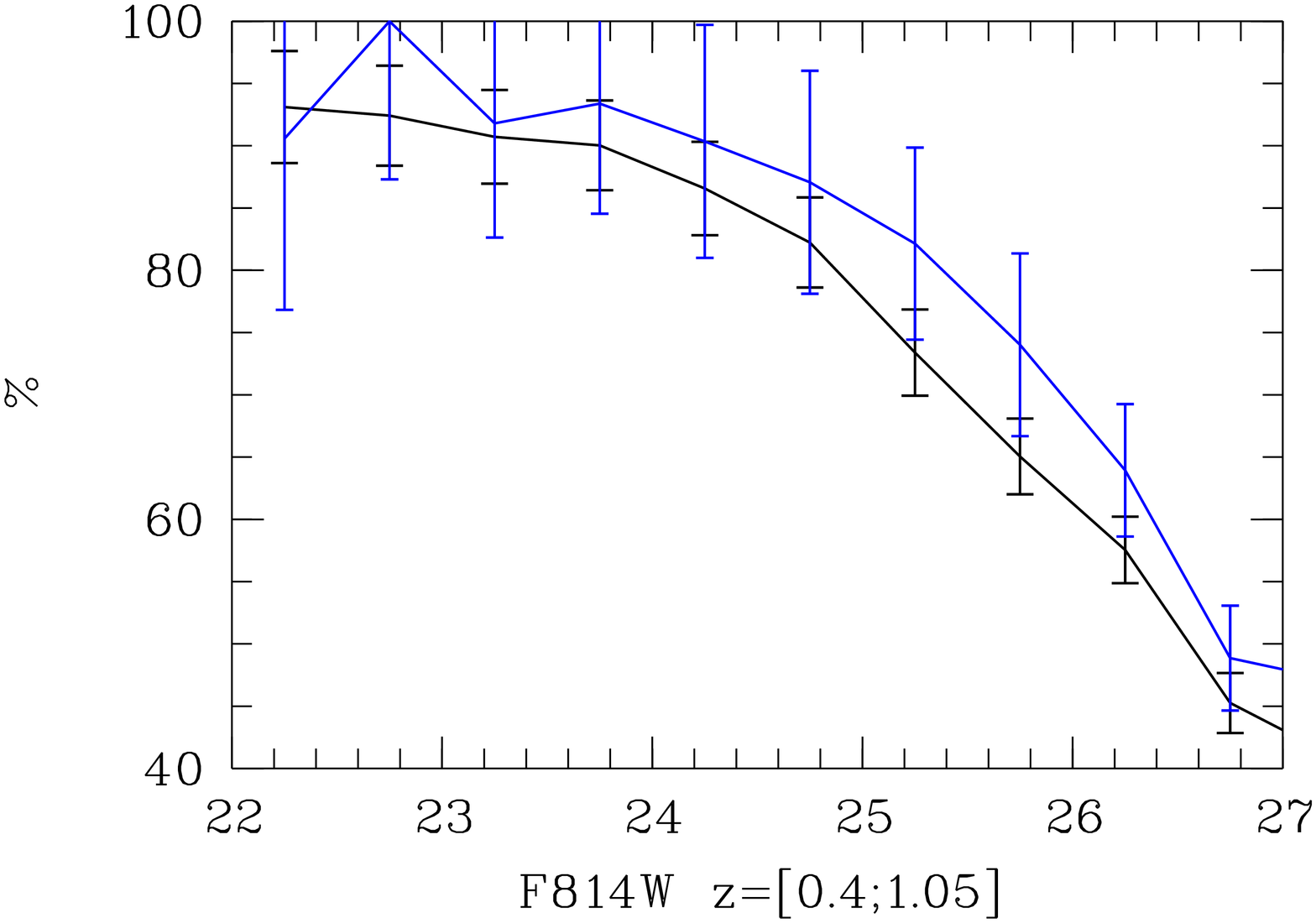}}
    \mbox{\includegraphics[angle=270,width=1.8in]{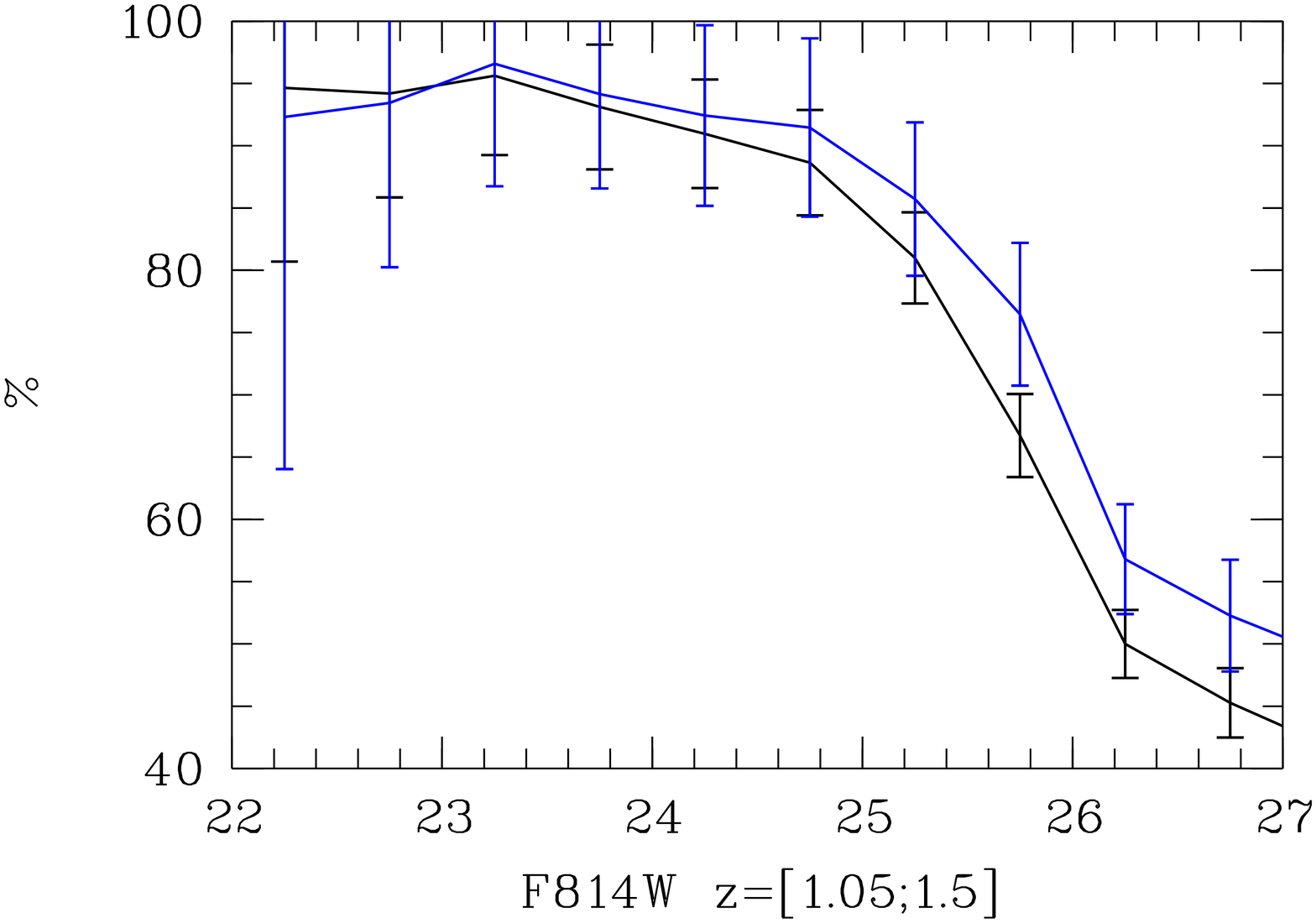}}
  }
  \centerline{
    \mbox{\includegraphics[angle=270,width=1.8in]{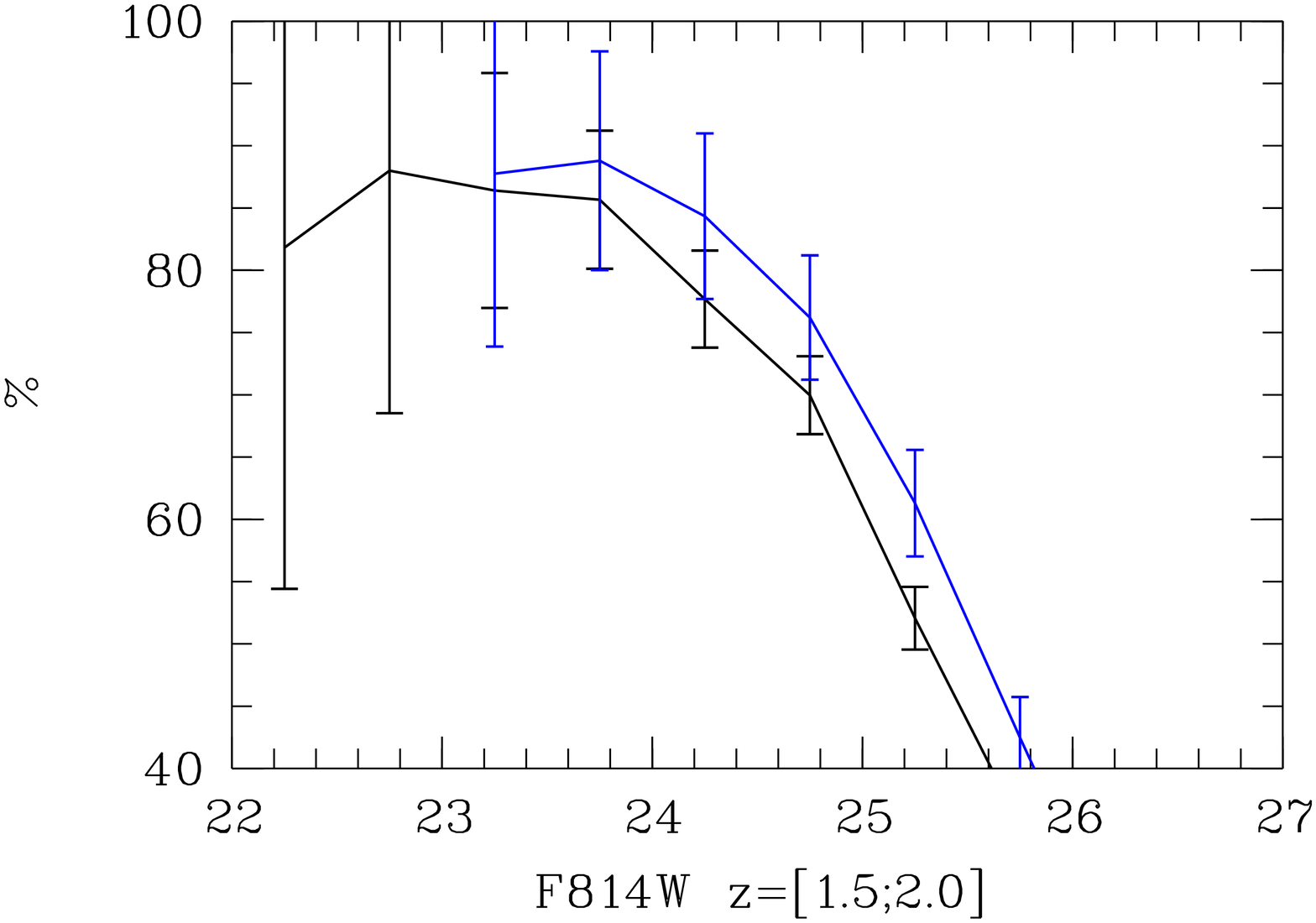}}
    \mbox{\includegraphics[angle=270,width=1.8in]{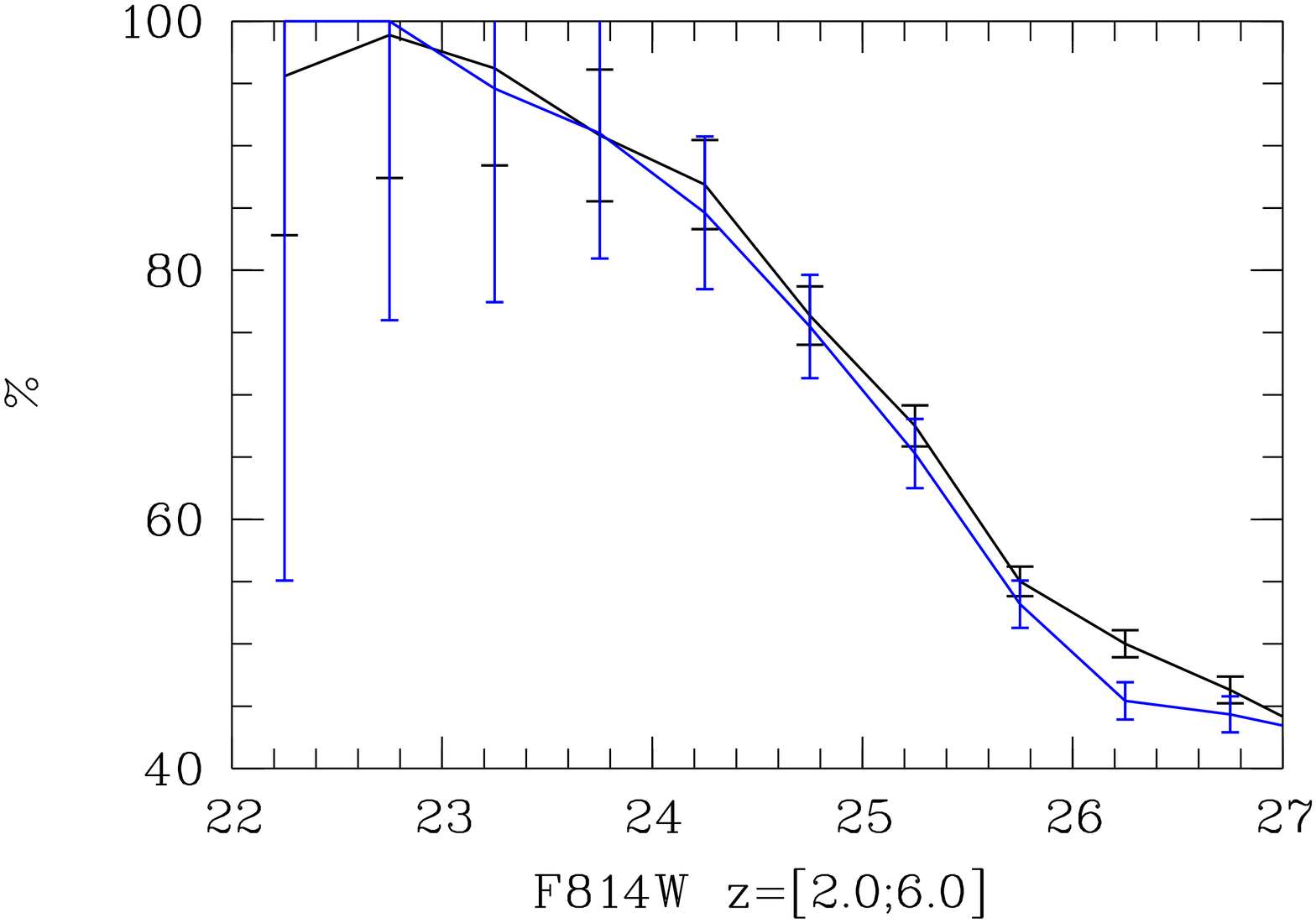}}
  }
  \caption{Percentages of galaxies with a \phz\ error lower than
    0.2*(1+z) as function of magnitude in four redshift bins. Black
    line: whole sample, blue line: galaxies polluted by a comparably
    bright or brighter galaxy. We computed the percentages only in the
    magnitude bins with more than 5 galaxies. }
  \label{histo}
  \end{figure}

We then computed the percentage of galaxies with a $1\sigma$ error bar
less than $0.2(1+z)$ (similarly as in Adami et al. 2008). This gives
us an estimate of the percentage of galaxies with \phzs\ which are not
catastrophic errors as a function of magnitude and as a function of
redshift, even beyond the spectroscopic limit. We plot in
Fig.~\ref{histo} these percentages for the ten merged lines of sight
considered. We first confirm that F814W$\leq$23.5 and z$\leq$1.05
galaxies probably have reliable \phzs, as suggested in the previous
section.  Moreover, percentages statistically remain globally higher
than 90$\%$ for magnitudes brighter than F814W$\sim$24 or 24.5 except
in the $z=[1.5;2.0]$ redshift range. As expected, the worse situation
appears for the $z=[1.5,2.0]$ range (see also Coupon et al. 2009). The
presently available magnitude passbands are not well adapted to this
range, the Balmer break being located redward of the z$'$ band, and
the Lyman break still being bluer than the B band. Finally, galaxy
fluxes contaminated by brighter close (blended) neighbors do not seem
to show values significantly different from isolated objects. Taken at
face value, these results suggest that our \phzs\ are not strongly
polluted by catastrophic errors down to F814W$\sim$24.5.

\begin{figure}
  \centerline{
    \mbox{\includegraphics[angle=270,width=1.8in]{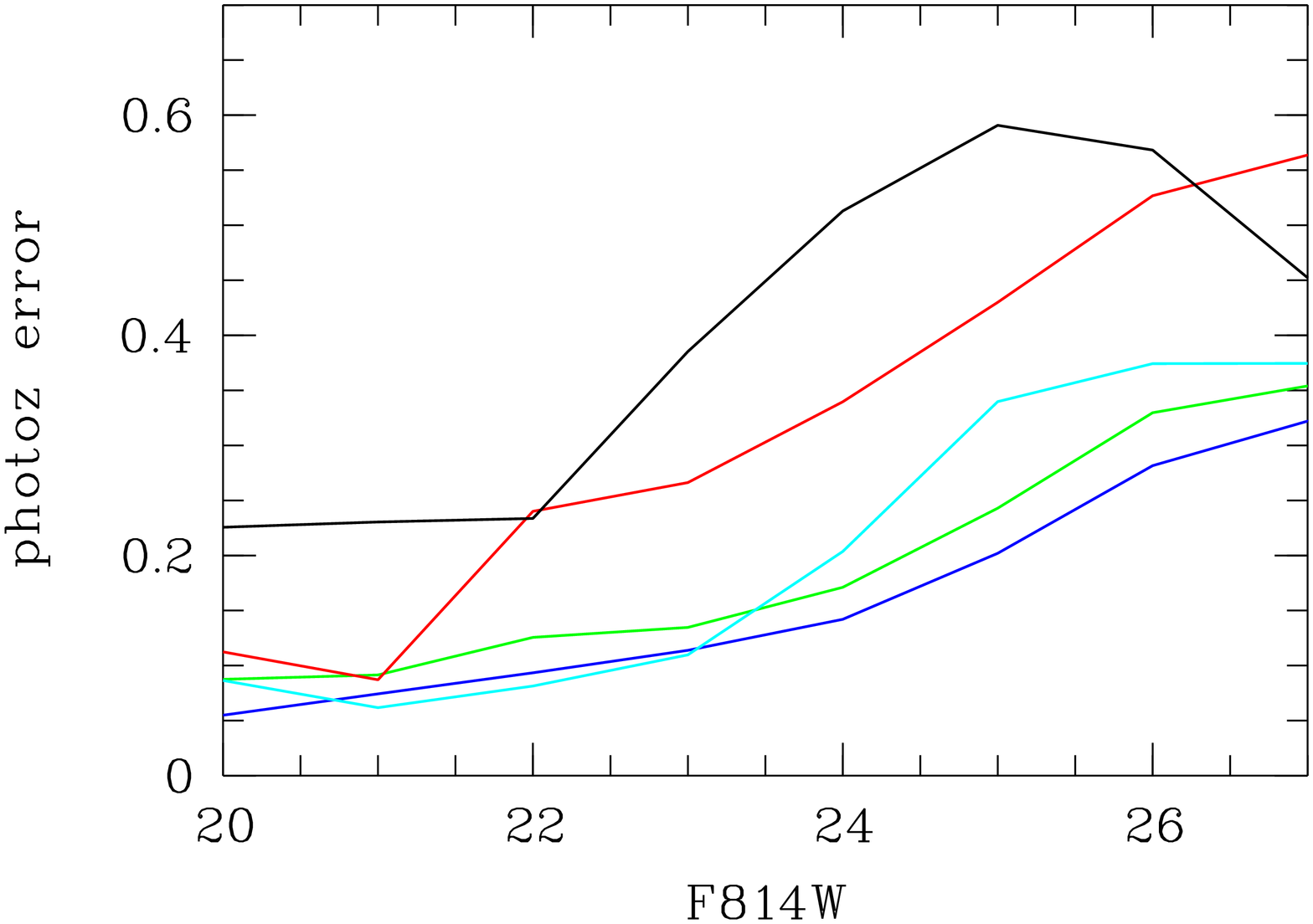}}
    \mbox{\includegraphics[angle=270,width=1.8in]{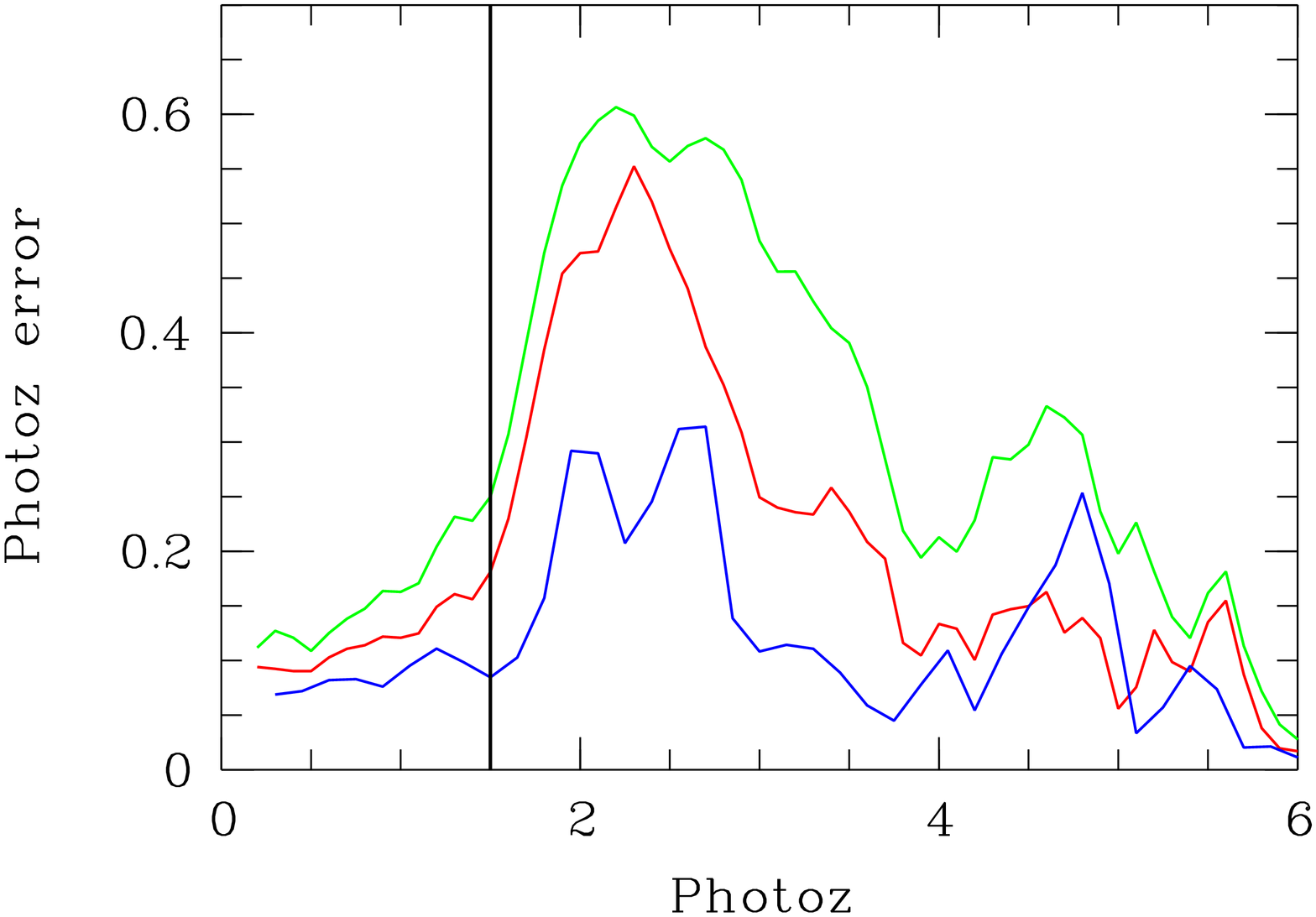}}
  }
  \caption{Left: mean individual \phz\ uncertainties (regular
    estimator) as a function of F814W magnitude in four redshift
    intervals, color coded as: blue: z=[0.4;1.05], green:
    z=[1.05;1.5], red: z=[1.5;2.0], black: z=[2.0;3.0], cyan: 
    z=[3.0;6.0]. Right: mean
    individual \phz\ uncertainties as a function of \phz\ for various
    F814W magnitude intervals, color coded as: blue: F814W=[19.5;22.5],
    red: F814W=[19.5;24.5], green: F814W=[19.5;26.]. The vertical line
    shows the z$\leq$1.5 limit we suggest to adopt in order to use the
    \phzs\ of this paper.}
  \label{meansig}
  \end{figure}

We plot in Fig.~\ref{meansig} the mean individual \phz\ errors per bin
of magnitude as a function of magnitude and \phz\ for various redshift
intervals. This gives us an estimate of the 1$\sigma$ uncertainty
around the $specz$/\phz\ relation in the considered redshift
interval. We confirm and extend the results of the previous
section. Galaxies brighter than F814W$\sim$24.5 and in the z=[0.4;1.5]
or brighter than F814W$\sim$24 and in the z=[3.0;6.0] range have relatively 
low \phz\ uncertainties. We also confirm that \phzs\ in the z=[1.5;3.75] 
range are poorly constrained.

These tests therefore lead us to adopt a conservative approach, and to
limit our catalogs to galaxies brighter than F814W=24.5 and
z$\leq$1.5. We may also consider galaxies brighter than F814W=24 and
at z$\geq$3.75.

\subsection{Photo-$z$ quality checks: the z$\geq$3 domain}

As an additional external test of the \phz\ uncertainties for distant
and faint galaxies, we took advantage of the giant arcs detected along
the LCDCS 0504 cluster (see Fig.~\ref{cl1216arc}). These arcs are
likely to be multiple images of a low number of sources and should
therefore have identical redshifts when they originate from a single
object.  We computed \phzs\ for these arcs and at least four of them
produced values close in redshift with similar spectral types (T in
[21,31], all consistent with an active galaxy). These arcs have F814W
magnitudes (measured in the rectangles shown in Fig.~\ref{cl1216arc})
of $26.3\pm 0.2$, $23.9 \pm 0.1$, $25.5 \pm 0.1$ and $24.8 \pm 0.1$.
We assumed that these four objects were multiple images of a single
z$\sim$3.68 galaxy, although we may have two couples of arcs coming
from two distinct objects at z$\sim$3.74 and z$\sim$3.62. This allowed
us to estimate an upper value of the \phz\ uncertainty in this
redshift range and at these magnitudes. We found a value of 0.07 even
better than the statistical estimates of Fig.~\ref{meansig}.

\begin{figure}
  \centerline{
%    \mbox{\includegraphics[angle=0,width=3in]{fig1216.ps}}
  }
  \caption{LCDCS 0504 F814W image showing the giant arcs for which we 
computed photometric redshifts.}
  \label{cl1216arc}
  \end{figure}

As a by product, we also performed  a dynamical analysis of
LCDCS 0504 from the giant detected arcs (see Appendix~A).

\subsection{Photo-$z$ quality and spatial resolution}

\begin{figure}[!h]
  \begin{center}
    \includegraphics[angle=270,width=3.00in]{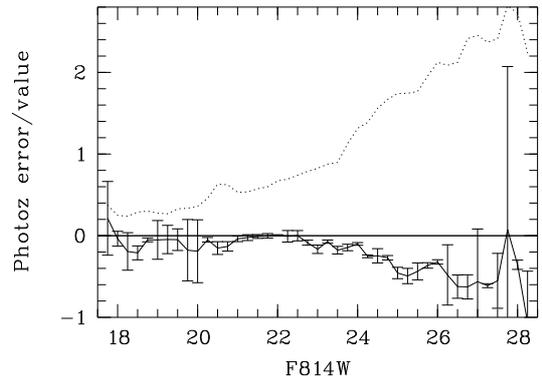}
    \caption{Continuous line (with error bars): \phz\ difference
      for LCDCS 0851 between degraded HST image and full HST
      resolution, as a function of magnitude. Dotted line: mean
      \phz\ values as a function of magnitude (full HST resolution).}
  \label{zz}
  \end{center}
  \end{figure}

We will now test if systematic effects appear as a function of image
resolution.  For this, we degraded the HST image of LCDCS0541 to a
1~arcsec resolution. As stressed above, this would lead to an
important loss of objects, so we concentrate here only on the \phz\
estimate. We repeat all the process of \phz\ computation (detection of
objects in the HST ACS F814W images, and measurements in the other
bands) and correlate the present \phz\ estimates with the values
coming from the image with no degradation. We show in Fig.~\ref{zz}
the differences between the two estimates, as a function of
magnitude. This figure provides evidence for a redshift underestimate
for F814W magnitudes fainter than 24. This underestimate is lower than
$\sim$0.2 down to F814W=24.5, which remains modest compared to the
mean \phz\ values also shown in Fig.~\ref{zz}. We therefore conclude
that the fact that the spatial resolutions of the considered images
differ is not a problem for \phzs\ down to F814W$\sim$24.5. This
conclusion is in good agreement with the previous sections, where we
also proposed to limit the use of \phzs\ to objects brighter than
F814W$\sim$24.5 (which corresponds to a mean redshift of 1.5, also in
good agreement with previous expectations).

\subsection{Do we need Spitzer images to estimate photo-$z$s?}

\begin{figure}[!h]
  \begin{center}
    \includegraphics[angle=270,width=3.00in]{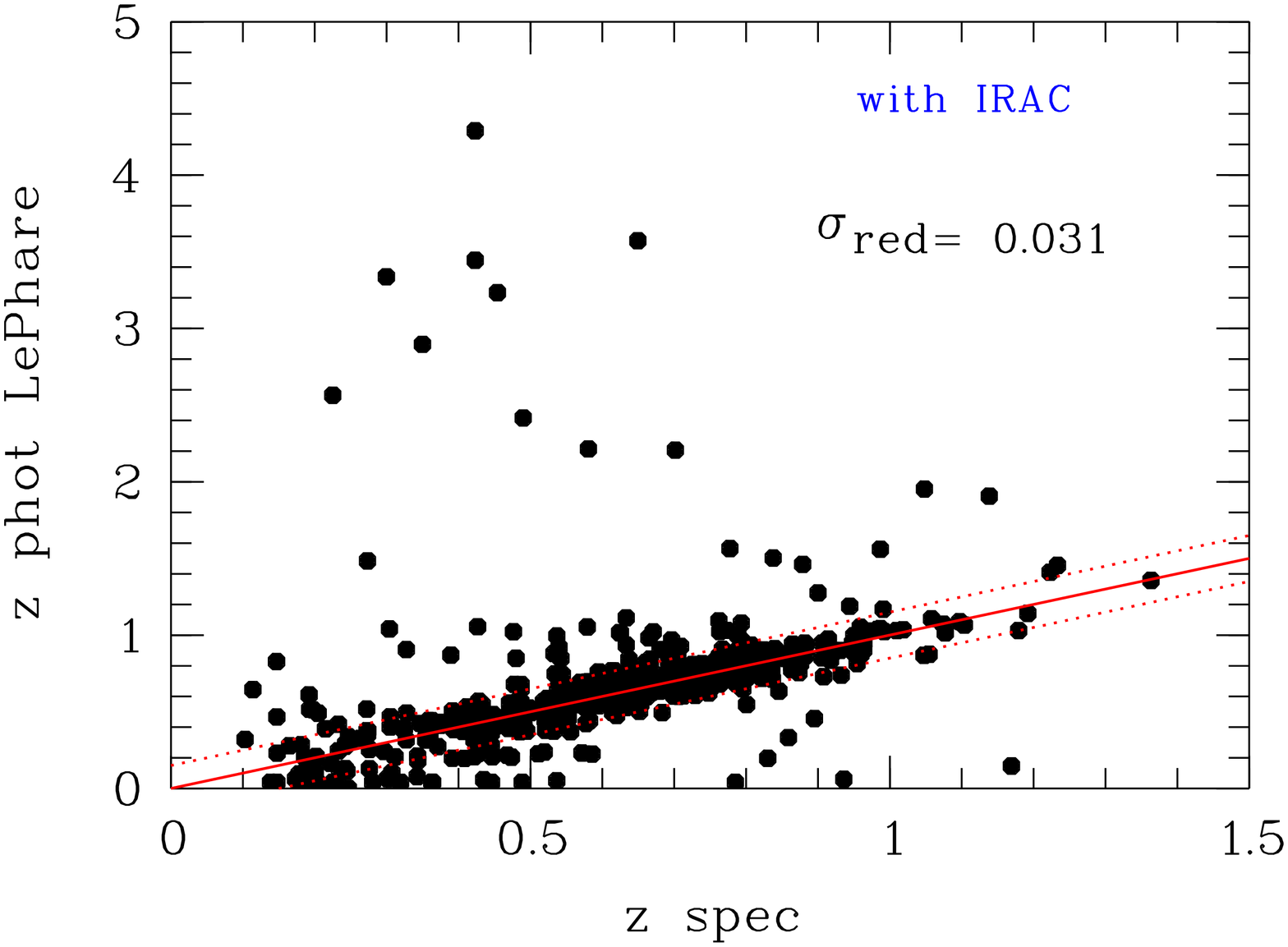}
    \includegraphics[angle=270,width=3.00in]{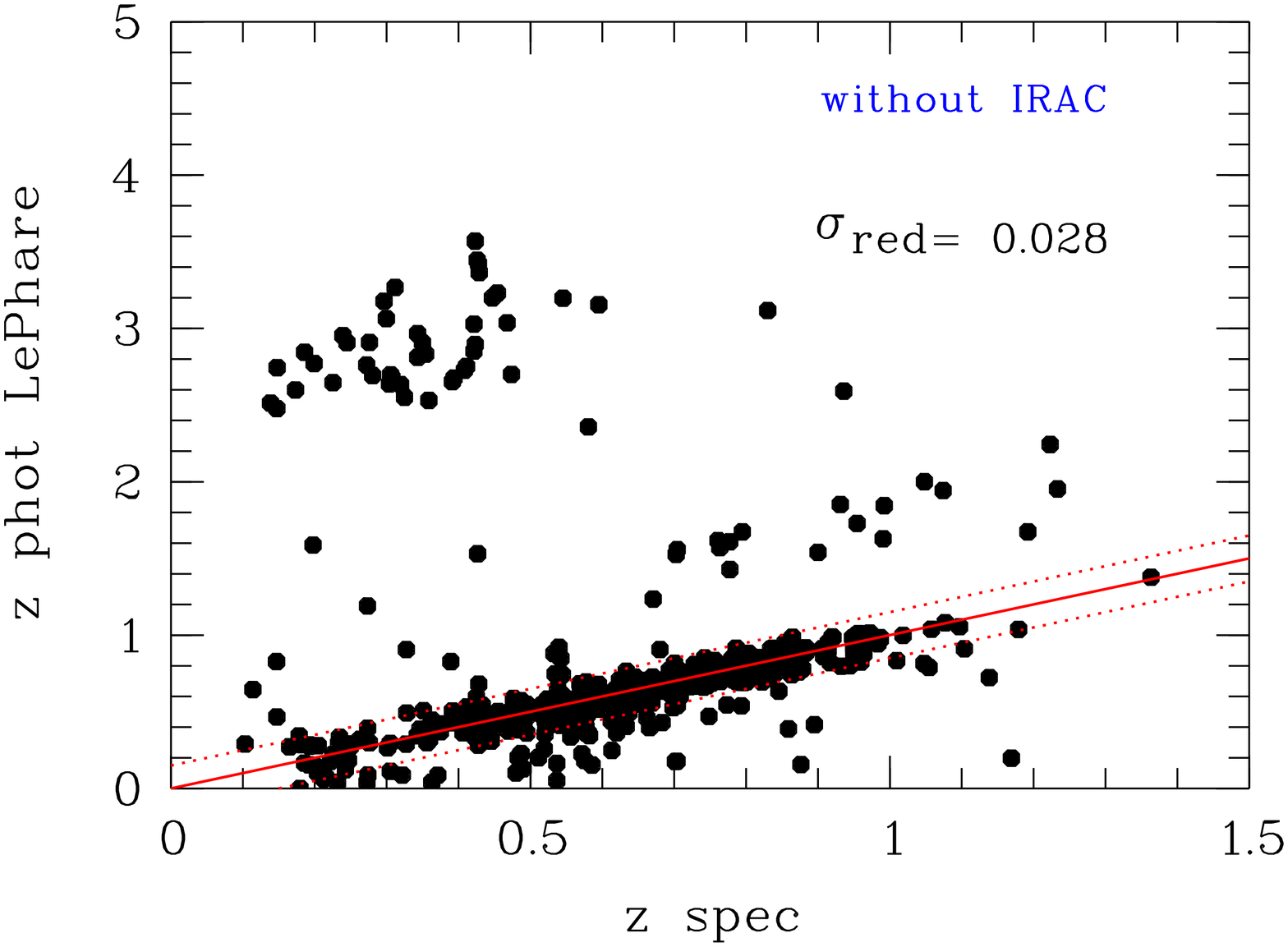}
    \includegraphics[angle=270,width=3.00in]{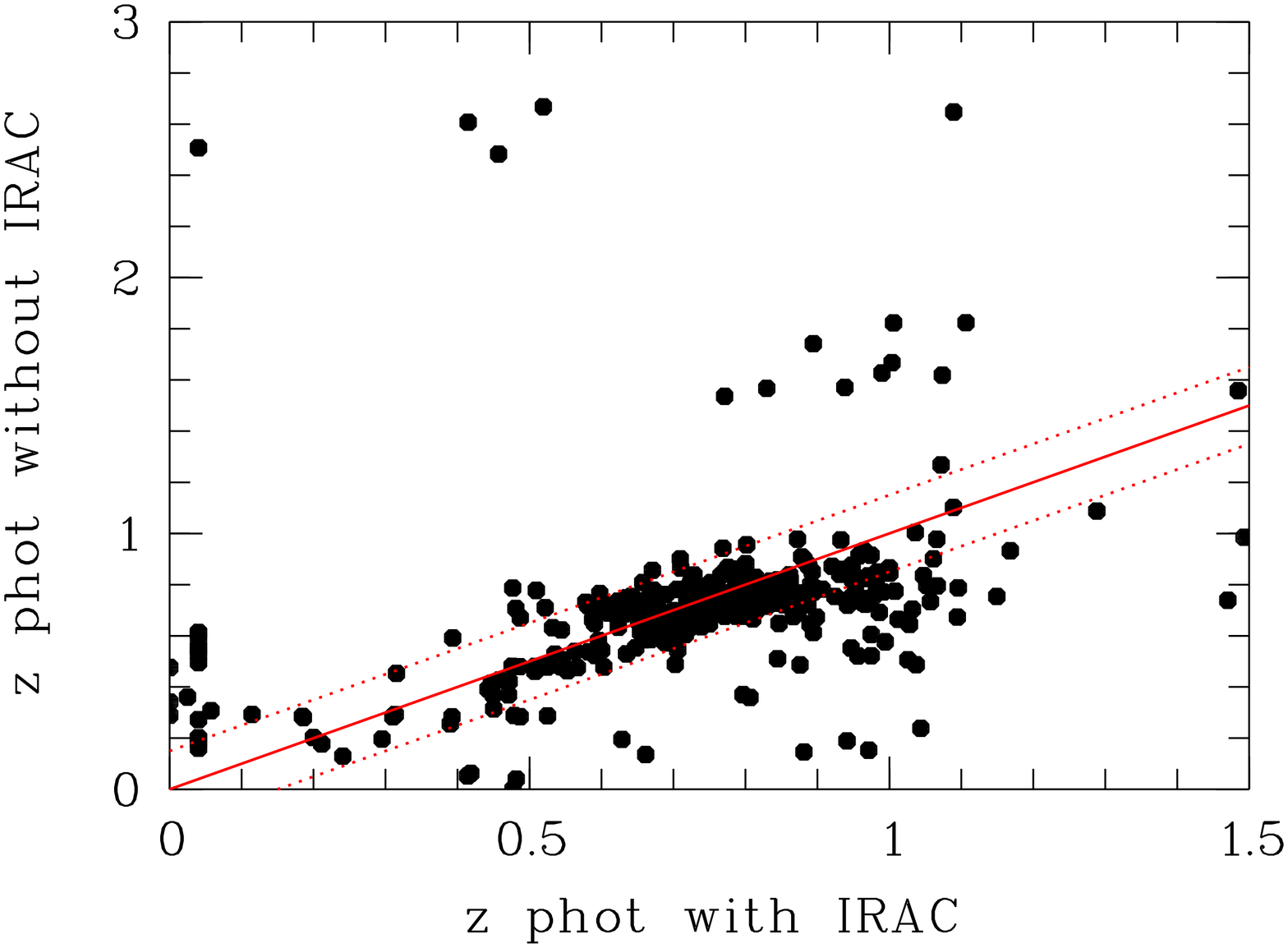}
    \caption{Spectroscopic versus photometric redshifts for the ten
      lines of sight. Top figure: \phzs\ computed with all
      available bands. Central figure: \phzs\ computed without
      Spitzer data. Bottom figure: \phzs\ computed with and without
      Spitzer data.}
  \label{wwS}
  \end{center}
  \end{figure}

The images with the worse spatial resolution in our sample are the two
Irac bands so we need to check if there is any benefit in using these
two bands, in particular for objects with F814W$\leq$23.5. We
therefore merged spectroscopic redshift catalogs for the ten lines of
sight considered and produced a $specz$ versus \phz\ diagram with and
without Spitzer data (see Fig.~\ref{wwS}). The reduced sigma around
the mean relation is better when Spitzer data are not considered (0.028
versus 0.031). However, the inclusion of these infrared data reduces by
a factor of 3 to 4 the catastrophic error percentage (2$\%$ versus 7$\%$)
as computed by Ilbert et al. (2006).
This catastrophic error percentage of 2$\%$ is comparable to
that found by Coupon et al. (2009).  It seems therefore
reasonable to pay the price of a slightly higher reduced sigma (but
still acceptable and comparable to classical literature values of 0.03
for LePhare, e.g. Coupon et al. 2009) to obtain a catastrophic error
percentage lower by a great factor. 

\section{Comparison with the literature}

\subsection{General redshift histogram}

Now that we have defined a well controlled sample, we compare in this
section the resulting \phz\ histogram (from the ten merged lines of
sight) with known literature redshift distributions.  We chose the
VVDS deep and ultra deep (LeF\`evre et al. 2005, LeF\`evre et al. 
in preparation) surveys
because they provide a much more robust spectroscopic redshift
distribution than any other literature \phz\ distribution. 
The VVDS deep survey gathers 7266
reliable spectroscopic redshifts down to I$_{AB}$$\sim$24.1 over a
$\geq$0.6 deg$^2$ area. The VVDS ultra deep survey gathers 550
reliable spectroscopic redshifts down to i'$_{AB}$$\sim$24.75 over a
more limited area of $\sim$400 arcmin$^2$. This magnitude limit is
very close to F814W=24.5 (an Sc galaxy at z=1.2 has i'$-$F814W$\sim$0.3)
and will allow a direct comparison without magnitude limitations.

\begin{figure}[!h]
\begin{center}
    \includegraphics[angle=270,width=3.50in]{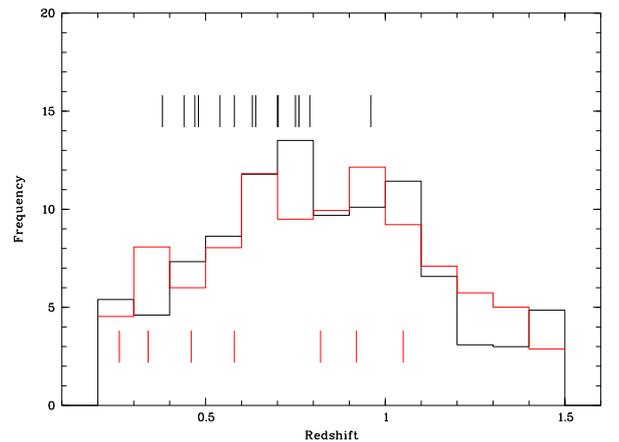}
    \caption{Normalized redshift distributions of our \phzs\ (black
      histogram) superimposed on the VVDS deep and ultra deep
      spectroscopic redshift distribution (red histogram). Short black
      vertical lines show the redshift distribution of known NED
      clusters along our lines of sight. Short red vertical lines show
      the redshift distribution of known clusters in the VVDS deep
      survey from Mazure et al. (2007).}
  \label{Nz}
  \end{center}
  \end{figure}

We re-normalized the two histograms (from the VVDS surveys and from
our \phz\ catalogs) to the same number of galaxies in order to take
into account the different spatial coverages. We then produced
Fig.~\ref{Nz} where we can see the generally good agreement between
the two distributions. Some differences are visible but are nearly all
explained by known clusters along the different lines of sight. The
only puzzling differences occur at z$\geq$1.2. They can be explained
by the fact that this redshift interval is mainly dominated by VVDS
ultra-deep data, which only cover a relatively small area of the sky
and are subject to a strong cosmic variance. A second explanation can
also be that \phzs\ begin to have a decreasing precision beyond
z$\sim$1.3 and F814W$\geq$22.5, as shown in Fig.~\ref{meansig}.

\subsection{Comparison with EDisCS photo-$z$s}

Although Pell\'o et al. (2009) already computed \phzs\ for the EDisCS galaxy
cluster sample, there are fundamental differences in
the present approach based largely on the difference in goals.

First, we did not use the same photometric bands, due to differing
scientific goals. Most of the \phzs\ of Pell\'o et al. (2009) were
computed with V, R, I, J, and Ks magnitudes (except for Cl1232-1250
where B, V, I, J, and Ks were selected) because the study was mainly
focused on z$\sim$[0.4, 0.8] cluster galaxy populations. All \phzs\ in
the present study were computed with B, V, R, I, F814W, Spitzer IRAC
3.6 $\mu$m, and 4.5$\mu$m (J and Ks data were not all publicly available
at the beginning of the study) because we are also interested in detecting
foreground z$\leq$0.4 galaxy populations (to avoid potential pollution
of distant galaxy samples) and very high redshift
objects. This should lead us to have a better precision at redshifts
lower than 0.4 (where the 4000~\AA\ break is not yet bracketed if the
B band is not used) and for very distant objects
(z$\geq$5). Conversely, our precision will be degraded at z$\geq$1.5,
where we lose the 4000~\AA\ break due to our lack of near infrared
data. The two approaches are therefore complementary.

Second, Pell\'o et al. used the 80's and 90's Coleman et al. (1980),
Bruzual $\&$ Charlot (1993), and Kinney et al. (1996) SEDs. We
considered the more recent Polletta et al. (2007) SEDs. These recent
templates are becoming standard SEDs because they were optimized for
near infrared and infrared data and are therefore well adapted to our
set of magnitudes. They were also selected for the COSMOS survey
(e.g. Ilbert et al. 2009).

\begin{figure}[!h]
  \begin{center}
    \includegraphics[angle=270,width=3.00in]{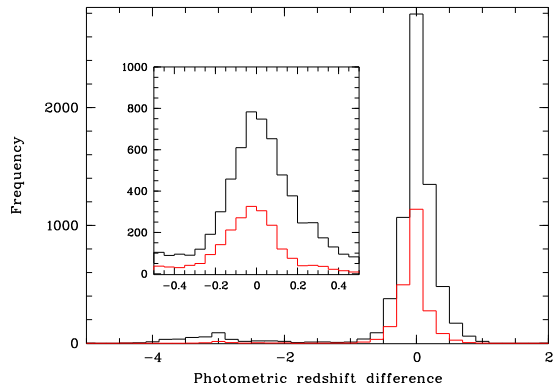}
    \caption{Histogram of the differences between the presently
      computed reliable \phzs\ (galaxies in 19.5$\leq$F814W$\leq$24.5,
      at redshift lower than 1.5, and with 1$\sigma$ LePhare
      uncertainty estimate lower than 0.2$\times$(1+z)) and the HyperZ
      \phzs\ of Pell\'o et al. (2009).  The black histogram is for
      F814W$\leq$24.5 and the red histogram is for
      F814W$\leq$22.5. The inner figure is a magnification of the
      central peak. }
  \label{zphotzphot}
  \end{center}
\end{figure}

For these two reasons, it is interesting to make a comparison between
the two \phz\ computations in the common redshift range z$\sim$[0.4,
1.5].  In this range, we compared our reliable \phzs\ (galaxies
brighter than F814W=24.5, and with 1$\sigma$ uncertainty lower than
0.2$\times$(1+z)) with the HyperZ values of Pell\'o et
al. (2009). Fig.~\ref{zphotzphot} shows a reasonable agreement. This
figure exhibits a few objects with large differences between the two
estimates. They are all at z$\ge$2 in the HyperZ computations, in good
agreement with the fact that Pell\'o et al.  (2009) \phzs\ were
computed with Ks data.  However they represent only 6$\%$ of our total
sample and are basically all fainter than F814W=22.5. This percentage
is also similar to the previously estimated percentages of
catastrophic errors in our estimates. The offset between the
two \phz\ estimates is $-0.03 \pm 0.20$ (1$\sigma$ error bar). This 
uncertainty of 0.20 has to be compared to the quadratic sum of the LePhare 
and HyperZ respective uncertainties.

%It is also interesting to compare the spectral types estimated with
%both methods. Removing catastrophic errors from the sample,
%Fig.~\ref{typetype} shows a linear relation between the two estimates:

%LePhare $T$ = ( 4.2 $\pm$ 0.2) HyperZ $T$ + (7.5 $\pm$ 0.8)

%The HyperZ types $T$ are E/S0 ($T$=1), Sbc ($T$=2), Scd ($T$=3), Im ($T$=4),
%Star-Bursts ($T$=5). They are therefore in relatively good agreement
%with our own estimates.

%\begin{figure}[!h]
%  \begin{center}
%    \includegraphics[angle=270,width=3.00in]{typetype.ps}
%    \caption{HyperZ versus LePhare estimated spectral types. The values
%quoted along the LePhare axis are the mean LePhare values for a given 
%HyperZ type. Error bars are $\pm 1\sigma$. We also plot individual values
%as small dots. We blured the HyperZ and LePhare integer types in 
%windows of width of respectively 1 and 5 in order to make the figure clearer.}
%  \label{typetype}
%  \end{center}
%  \end{figure}

These comparisons therefore validate our current approach.

\section{Summary}

We computed \phzs\ with the LePhare package for ten relatively distant
cluster lines of sight selecting B, V, R, I, F814W, z', Spitzer IRAC
3.6 $\mu$m, and 4.5$\mu$m images. These images were reduced and
aligned at the pixel scale using the SCAMP and SWarp tools. The zero
points of the various bands were adjusted by LePhare using publicly
available spectroscopy.

  The \phzs\ prove to be reliable in the z$\sim$[0.4, 1.5] redshift
  range and in the magnitude range F814W$\sim$[19.5, 24.5]. They are
  also relatively reliable in the z$\sim$[3.75, 6.0] redshift range
  and in the magnitude range F814W$\sim$[19.5, 24.].  We remarked that
  catastrophic errors mainly occured towards the high photometric
  redshifts (at z$\geq$1.5). This will obviously not affect our survey
  when limiting our analysis to the [0.4, 1.5] redshift range. The
  only consequence would be to remove a small number of galaxies. If
  we also consider the z$\sim$[3.75, 6.0] redshift range, we
  will include in our future weak lensing analyses some galaxies (of the order 
  of 2$\%$ from the spectroscopic redshift sample estimate) with
  completely wrong redshifts. Given the limited amount of such
  galaxies, the consequences on our survey will however remain
  limited.  We achieved a \phz\ precision of the order of 0.05 for the
full sample. This precision is degraded by a factor of two when
considering blended objects.

The \phz\ catalogs produced will therefore allow future weak lensing
tomography measurements as well as mass modeling of the 10 clusters
presently considered. We already used strong lensing features in
  order to model the mass of LCDCS 0504 (see Appendix~A).

  We also present evidence for environmental dependence of the
  \phz\ precision. We show that in dense regions (cluster centers or
  peripheral cluster areas), \phz\ precision for bright galaxies is of
  the order of 0.07 while it is of the order of 0.04 for faint
  galaxies. We detect the same variation in clusters when considering
  galaxies as a function of their spectral type. The \phzs\ are better
  estimated for S0 galaxies ($\sim$0.04) than for late type galaxies
  by a factor of about 2. Considering now galaxies in the sample not
  related to clusters, the \phz\ precision only slightly varies
  as a function of galaxy spectral type and absolute magnitude.  We
  reach similar conclusions in Adami et al. (2010) based on Coleman et
  al. (1980) SEDs instead of the Polletta et al. (2007) SEDs.  The
  catastrophic error percentages give similar results.  The largest
  catastrophic error percentages occur in clusters for bright and late
  type galaxies. In the field, we have similar percentages whatever
  the galaxy magnitudes and spectral types.  This leads us to conclude
  that regular field-based SEDs available in the literature are not
  very well adapted to high density environments. The agreement
  between spectroscopic and photometric redshifts stays acceptable
  most of the time, but could penalize cluster studies when a precise
  cluster membership is required. In this framework, we therefore plan
  in a future work to build such high-density region SEDs with long
  based spectral range instruments as VLT/X-Shooter, in order to have
  SEDs usable for cluster galaxies. The
  VLT/X-Shooter spectral coverage of $\sim$[300,2500]nm for z$\sim$0 galaxies 
  should for example be able
  to provide a large enough interval to constrain B to Irac1 bands at
  z$\sim$0.4 and V to Irac2 bands at z$\sim$0.9. We also plan to observe
  more distant objects than z$\sim$0 galaxies in order to limit possible 
  evolutionary effects. The most important
  galaxies to target are bright objects (M$_{i'}\leq-23$) making these 
  targets easy to observe. The
  VLT/X-Shooter should for example be able to provide signal to noise better
  than 5 over the available spectral range in a 2 hours integration for a 
  z=0.1 elliptical galaxy and better than 5 redder than 4000 A in a 4 hours 
  integration for a z=0.4 elliptical galaxy.
 
As external tests, we also compared the redshift histogram of our
\phzs\ with the redshift histogram from a spectroscopic survey, the
VVDS, and we find a good agreement. We also directly compared our
\phzs\ with the estimates of Pell\'o et al. (2009), and we also have a
good agreement in the redshift range where both computations are
reliable.

Finally, we stress that the data we present here are part of a larger
survey, the DAFT/FADA survey.  The present paper will therefore act as a 
reference
study for all subsequent articles. An important aspect is
that our data will be made available to the astronomical community at
the end of the survey, via a dedicated structure included in the
Marseille CENCOS data center (http://cencosw.oamp.fr/).  

\begin{acknowledgements}
  The authors thank the referee for useful and detailed comments.
  This work was supported in part by Department of Energy grant number
  DE-FG02-08ER41567 and in part by the French PNCG. We thank
  R. Pell\'o for providing us with the EDisCS photometric
  redshifts. We also thank O. LeF\`evre and the VVDS team for giving
  us their redshift distribution prior to publication. DC acknowledges 
  support from the Alfred P. Sloan Foundation. Last but not
  least, we are very grateful to E.~Bertin for his help in helping us
  to use his softwares.

\end{acknowledgements}

\appendix

\section{The Einstein Radius of LCDCS 0504}

  LCDCS 0504 has prominent arc features (see Fig.~\ref{cl1216arc})
  that form a near circular ring. We can use these arcs to make a
  simple estimate of the mass inside the ring by assuming circular
   symmetry. A completely circular mass model results in a
  complete Einstein ring. We calculate the Einstein radius for the
  three knots in this arc (seen on both sides of the cluster: a, b, c,
  see Fig.~\ref{cl1216arcbis}) and find $\theta_E = 16.28 \pm 0.02$
  arcseconds.

\begin{figure}[!h]
  \begin{center}
    \caption{LCDCS 0504 trichromic image showing the arcs selected to estimate
the Einstein radius.}
  \label{cl1216arcbis}
  \end{center}
  \end{figure}

The Einstein radius for a circular mass distribution is defined by the equation:
\begin{equation}
\theta_E = \sqrt{\frac{4 G ~m(\theta_E)}{c^2}~ \frac{d_{LS}}{d_{L}d_{S}}}
\end{equation}
where $m(\theta)$ is the projected or cylindrical mass inside an angle $\theta$
and $d_L$, $d_S$ and $d_{LS}$ are the usual angular diameter distances, to the lens, to 
the source and between the lens and the source respectively. It is easily shown that 
this equation is equivalent to the equation $\gamma(\theta_E)+\kappa(\theta_E)$=1 
where $\gamma$ and $\kappa$ are the shear and convergence. 

For an isothermal sphere, with velocity dispersion $\sigma$,
$\gamma(\theta)=\kappa(\theta) = 2 \pi (\sigma/c)^2 (d_{LS}/d_{S})
~\theta^{-1}$ and so $\theta_E = 4 \pi (\sigma/c)^2 (d_{LS}/d_{S})$.

We have measured the photometric redshifts of the faint blue arcs and
determined $z_S=3.7$.  The redshift of the lens is $z_L=0.7943$
(Halliday et al. 2004). This allows us to calculate the angular
diameter distances and to determine the velocity dispersion of the
isothermal model. We find, $\sigma=969 \pm 50$ km/s with the 5\% error
dominated by the model dependence and the assumption of circular
symmetry. This is in good agreement with $\sigma= 1018 \pm 75$ km/s,
the velocity dispersion measured by Halliday et al. (2004).

Another model to consider would be the NFW model. The Einstein radius
can be calculated numerically from the condition
$\gamma(\theta_E)+\kappa(\theta_E)$=1 using the well known formulae
(e.g. Wright \& Brainerd 2000) for the NFW profile. The NFW model has
two parameters $r_{200}$ and $c_{200}$. With the one constraint from
$\theta_E$, there is a degeneracy between these two parameters and so
$r_{200}$ depends on the assumption on $c_{200}$ and vice
versa. We give the resultant $r_{200}$ and $M_{200}$ (the 3D mass
inside $r_{200}$) for various $c_{200}$ in Table~\ref{NFW}.

\begin{table}
\caption{$r_{200}$ and $M_{200}$ (the 3D mass inside $r_{200}$) for various $c_{200}$.}
\label{NFW}
\begin{center}
\begin{tabular}{ccc}
\hline
      $c_{200}$   &  $R_{200} ~(h^{-1}$ Mpc)   &  $M_{200} ~(h^{-1} M_{\odot})$ \\
\hline
      3   &   1.19 &  9.00 $\times 10^{14}$\\
      5   &   0.98 &  5.62 $\times 10^{14}$\\
      7   &   0.87 &  3.96 $\times 10^{14}$\\
      10  &   0.79 &  2.61 $\times 10^{14}$ \\
\hline
\end{tabular}
\end{center}
\end{table}

In general the relation between $c_{200}$ and $r_{200}$, consistent
with our constraint, is given by the following fitting function:

$ln(r_{200}) = \sum_i p_i \ln(c_{200})^i$ with $p_0=0.58485295, p_1= -0.29387216, p_2=-0.12836863, p_3=0.057980625 $ and $p_4=-0.0061749105 $.

\end{document}